\documentclass[aps,prd,twocolumn,superscriptaddress,nofootinbib]{revtex4-2}

\usepackage{graphicx}
\usepackage{xcolor}
\usepackage{soul}
\usepackage{amsmath}
\usepackage[hidelinks]{hyperref}
\usepackage[normalem]{ulem}

\newcommand{\orcid}[1]{\href{https://orcid.org/#1}{\includegraphics[width=9pt]{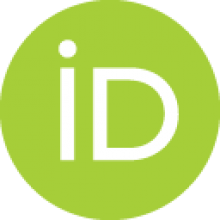}}}

\begin{document}

\title{The Fast Flavor Instability in Hypermassive Neutron Star Disk Outflows}

\author{Rodrigo Fern\'andez \orcid{0000-0003-4619-339X}}
\email{rafernan@ualberta.ca}
\affiliation{Department of Physics, University of Alberta, Edmonton, AB T6G 3E1, Canada}

\author{Sherwood Richers \orcid{0000-0001-5031-6829}}
\affiliation{Department of Physics, University of California, Berkeley, CA 94720, USA}

\author{Nicole Mulyk \orcid{0000-0001-7491-046X}}
\affiliation{Department of Physics, University of Alberta, Edmonton, AB T6G 3E1, Canada}

\author{Steven Fahlman \orcid{0000-0003-4875-9940}}
\affiliation{Department of Physics, University of Alberta, Edmonton, AB T6G 3E1, Canada}

\date{\today}

\begin{abstract}
We examine the effect of neutrino flavor transformation by the fast
flavor instability (FFI) on long-term mass ejection from accretion disks formed
after neutron star mergers. Neutrino emission and absorption in the disk set
the composition of the disk ejecta, which subsequently undergoes $r$-process
nucleosynthesis upon expansion and cooling.  Here we perform 28 time-dependent,
axisymmetric, viscous-hydrodynamic simulations of accretion disks around
hypermassive neutron stars (HMNSs) of variable lifetime, using a 3-species
neutrino leakage scheme for emission and an annular-lightbulb scheme for
absorption.  We include neutrino flavor transformation due the FFI in a
parametric way, by modifying the absorbed neutrino fluxes and temperatures,
allowing for flavor mixing at various levels of flavor equilibration, and also in
a way that aims to respect the lepton-number preserving symmetry of the
neutrino self-interaction Hamiltonian.  We find that for a promptly-formed
black hole (BH), the FFI lowers the average electron fraction of the disk
outflow due to a decrease in neutrino absorption, driven primarily by a drop in
electron neutrino/antineutrino flux upon flavor mixing.  For a long-lived HMNS,
the disk emits more heavy lepton neutrinos and reabsorbs more electron neutrinos than for a BH, with a
smaller drop in flux compensated by a
higher neutrino temperature upon flavor mixing. The resulting outflow has a broader 
electron fraction distribution, a more proton-rich peak, and undergoes stronger
radiative driving. Disks with intermediate HMNS lifetimes show results that
fall in between these two limits. In most cases, the impact of the FFI on the
outflow is moderate, with changes in mass ejection, average velocity, and
average electron fraction of order $\sim 10\%$, and changes in the
lanthanide/actinide mass fraction of up to a factor $\sim 2$.
\end{abstract}

\maketitle

%%%%%%%%%%%%%%%%%%%%%%%%%%%%%%%%%%%%%%%%%%%%%%%%%%%%%%%%%%%%%%%%%%%%%%%
\section{Introduction}

Neutron star (NS) mergers became the first confirmed cosmic site of $r$-process
element production, following the detection of a kilonova from GW170817
\cite{ligogw170817multi-messenger,coulter_2017,Cowperthwaite2017,drout2017,Tanaka2017,tanvir2017}.
Nucleosynthesis takes place in the expanding ejecta, which is neutron-rich and
therefore favors the operation of the $r$-process
\cite{Lattimer1974,Eichler+89,Freiburghaus+99}. The ejecta is made up of
multiple components launched over a range of timescales and by a variety of
mechanisms (e.g.,
\cite{FM16,baiotti_BinaryNeutronStar_2017,radice_DynamicsBinaryNeutron_2020}).
Of particular importance is matter unbound from the accretion disk formed after
the merger, which can dominate mass ejection in events like GW170817 (e.g.,
\cite{shibata_2017b}).

Transport of energy and lepton number by neutrinos is a key physical process in
the accretion disk, because the timescales associated with some of the ejection
mechanisms are comparable to or longer than the weak interaction timescale.
Neutrino transport heats or cools different parts of the disk and modifies the
electron fraction of the disk material (e.g, \cite{Ruffert+97}).  Neutrinos can
also be involved in the launching of a gamma-ray burst jet, by clearing out
dense matter from the polar regions, or contributing energy through
neutrino-antineutrino pair annihilation (e.g.,
\cite{goodman_1987,Eichler+89,Richers2015,just_2016,foucart_2018,fujibayashi_PropertiesNeutrinodrivenEjecta_2017}).
Given the thermodynamic conditions reached in NS mergers, however, temperatures
are well below the muon and taon mass energies ($\sim 100$\,MeV and $\sim 1.8$\,GeV, respectively),
thus electron-type neutrinos and antineutrinos are the only species
that can exchange energy and lepton number with matter locally or non-locally
through charged current weak interactions, with heavy lepton neutrinos
fulfilling primarily a cooling role\footnote{The exception being neutrino pair
annihilation in low-density polar regions (e.g., \cite{sumiyoshi_2021}).}.

Flavor transformation due to non-zero neutrino mass and to interactions
with background matter (the MSW mechanism
\cite{mikheyev_ResonantNeutrinoOscillations_1989,wolfenstein_NeutrinoOscillationsMatter_1978})
have long been expected to occur at large distances from the merger, with
little impact on the dynamics or nucleosynthesis of the ejecta. However, neutrino-neutrino
interactions make the flavor transformation process nonlinear, leading to a
rich phenomenology (e.g,
\cite{duan_2010_arnps,mirizzi_SupernovaNeutrinosProduction_2016,capozzi_NeutrinoFlavorConversions_2022}).
In the context of neutron star mergers, the matter-neutrino resonance was shown
to occur in the polar regions above the remnant, such that it could have significant impacts on
the nucleosynthesis in outflows along the rotation axis
\cite{malkus_2016,wu_msw_2016}. More recently, the so-called \emph{fast flavor
instability} (FFI) was shown to be ubiquitous in both neutron star mergers and
core-collapse supernovae, resulting in extremely fast (nanosecond) flavor
transformation both within and outside of the massive accretion disk
\cite{wu_tamborra_2017} and the HMNS \cite{george_2020}.

Although local simulations of the FFI have been performed and can predict the
final flavor abundance following the instability
\cite{bhattacharyya_2020,padilla_2021,wu_2021_1d,richers_2021_pic,richers_2021_3d,duan_FlavorIsospinWaves_2021,martin_FastFlavorOscillations_2021,zaizen_NonlinearEvolutionFast_2021,xiong_PotentialImpactFast_2020,sigl_SimulationsFastNeutrino_2022}
(see also \cite{richers_code_comp} for a code comparison study), a general
description of the effects of the instability and a consistent inclusion in
global simulations is still lacking. 
Assessment of the FFI in post-processing of time-dependent
simulations of NS merger remnants has confirmed the prevalence of the
instability outside the neutrino decoupling regions, with implications for the
composition of the disk outflow
\citep{wu_2017_trajectories,george_2020,richers_EvaluatingApproximateFlavor_2022}.

Effective inclusion of the FFI in global simulations of
post-merger black hole (BH) accretion disks has been achieved recently, first in 
general-relativistic (GR) magnetohydrodynamic (MHD) 
simulations over a timescale of $400$\,ms \cite{Li_Siegel_2021}, and
then also on viscous hydrodynamic simulations over the full 
evolutionary time of an axisymmetric disk ($\sim 10$\,s, \cite{Just2022_FFI}, who
also performed 3D MHD simulations for $500$\,ms).
In both cases, a standard 3-species, 2-moment scheme (M1) was
modified based on a criterion indicating fast flavor instability,
along with an algebraic swapping scheme between species to mix the zeroth and first
moments. Both studies found that the FFI results in a 
$\sim 10\%$ decrease in mass ejection, with the ejecta shifting toward
more neutron-rich values. The prevalence of the instability over the
entire disk system was confirmed, and the sensitivity to various
mixing prescriptions was found by \cite{Just2022_FFI} to be moderate.

Here we introduce a different method to include the effects of the FFI in
global simulations that employ a leakage-lightbulb-type neutrino scheme, in
order to enable parameter studies over a larger number of long-duration
accretion disk simulations. We employ an optical depth prescription to smoothly
activate the FFI in regions where neutrinos are out of thermal equilibrium, and
use algebraic expressions to parametrically mix the fluxes and energies of each
neutrino flavor absorbed by the fluid. The scheme relies on the very rapid
growth and saturation of the instability ($\sim $\,ns timescales over $\sim
$\,cm length scales) relative to the relevant evolutionary time- and spatial
scales of the system ($> $\,ms timescales over $\sim$\,km length scales). The
efficiency of our method allows for exploration of varying degrees of flavor
equilibration, as well as flavor mixing that respects lepton number
conservation in the neutrino self-interaction Hamiltonian. 

We apply this method self-consistently to an axisymmetric viscous hydrodynamic
setup representative of a post-merger accretion disk, and explore the effects
of the instability on long-term mass ejection from disks around hypermassive
neutron stars (HMNSs) of variable lifetime. Viscous hydrodynamic simulations
that include neutrino emission and absorption as well as nuclear recombination
produce ejecta that is consistent with GRMHD simulations at late-time
($\gtrsim$\,s timescales), since viscous heating models dissipation of MHD
turbulence reasonably well, with the main difference being the lack of earlier
ejecta launched by magnetic stresses \cite{F19_grmhd}.  Thus, our results
produce a lower limit to the quantity of ejecta from these systems, while also
focusing on the portion of the ejecta that is most affected by neutrinos.

The paper is structured as follows. Section \S\ref{s:methods} describes
the hydrodynamics simulations, the neutrino implementation, flavor
transformation prescription, and models evolved. Results are presented
in \S\ref{s:results}, including evolution without and with flavor transformation,
nucleosynthesis implications, and comparison with previous work. A summary
and discussion follow in \S\ref{s:summary}. Appendix~\ref{s:asymmetric} provides
a derivation of our lepton-number-preserving prescription for FFI flavor transformation.

%%%%%%%%%%%%%%%%%%%%%%%%%%%%%%%%%%%%%%%%%%%%%%%%%%%%%%%%%%%%%%%%%%%%%%%
\section{Methods \label{s:methods}}

%----------------------------------------------------------------------------------
\subsection{Numerical Hydrodynamics \label{s:hydro}}

We solve the equations of time-dependent hydrodynamics in axisymmetry using
{\tt FLASH} version 3.2 \cite{fryxell00,dubey2009}, with the modifications
described in \cite{FM13,MF14,FKMQ14,lippuner_2017}. The code solves the
equations of mass, momentum, energy, and lepton number conservation in
spherical polar coordinates $(r,\theta)$, subject to the pseudo-Newtonian
potential of a spinning BH \cite{artemova1996} with no self-gravity, an
azimuthal viscous stress with viscosity coefficient $\alpha_{\rm v}$
\cite{shakura1973}, and the equation of state of \cite{timmes2000} with the
abundances of neutrons, protons, and alpha particles in nuclear statistical
equilibrium, accounting for nuclear binding energy changes.

Neutrino effects in the disk are included through a leakage scheme for cooling
and annular lightbulb irradiation with optical depth corrections for absorption
\cite{FM13,MF14,lippuner_2017}. The HMNS is modeled as a reflecting inner
radial boundary from which additional neutrino luminosities are imposed.  In
\S\ref{s:leakage} we describe the baseline neutrino scheme, including upgrades
relative to versions used in previous work, and modifications to include flavor
transformation due to the FFI.

The initial condition is an equilibrium torus with constant angular momentum,
entropy, and composition \cite{FM13}. The disk configuration and central object
mass is the same in all the simulations, aiming to match the parameters of
GW170817 (c.f., \cite{fahlman_2018}) and to connect with previous long-term
post-merger disk calculations (e.g., \cite{fujibayashi2018,Just2022_FFI}).  The
central object has a mass $2.65M_\odot$, spin $0.8$ if a BH, or otherwise a
radius $30$\,km and rotation period\footnote{The rotation 
period of the HMNS affects the way in which the viscous stress is applied at the
surface, where rigid rotation is enforced. The pseudo-Newtonian potential is
set to have spin zero when the HMNS is present.} $3$\,ms if a HMNS. 
The disk has a mass
$0.1M_\odot$, radius of maximum density $r_{\rm d}=50$\,km,
initial $Y_e = 0.1$, and entropy $8$\,k$_{\rm B}$ per baryon. The
viscosity parameter in all simulations is $\alpha_{\rm v}=0.03$.  The computational
domain outside the torus is filled with an inert low-density ambient medium.
The initial ambient level and density floors are set as described in
\cite{Fernandez2020BHNS}.

The computational domain spans the range $\theta \in [0,\pi]$ in polar angle,
with reflecting boundary conditions at each end of the interval. The domain is
discretized with a grid equispaced in $\cos\theta$ using $112$ cells.  In the
radial direction, the grid is logarithmically spaced with $128$ points per
decade in radius. This results in a resolution $\Delta r / r \simeq \Delta
\theta \simeq 0.02$ at the equator.  When a BH sits at the center, the inner
radial boundary is set at $r\simeq 8.8$\,km, halfway between the ISCO and the horizon
of the BH, and the boundary condition is set to outflow. When a HMNS is
present, the inner radial boundary is reflecting and set at $r=30$\,km. The
outer radial boundary is a factor $10^5$ times larger than the inner radial
boundary, and the boundary condition is set to outflow.

When a HMNS is transformed into a BH, the inner radial boundary is moved inward
(from $30$\,km to $8.8$\,km), the extension to the computational domain is
filled with values equal to the first active cell outside the HMNS prior to
collapse, the imposed HMNS luminosities are turned off, and the inner radial
boundary is set to outflow.  The newly added cells are filled with inert
matter: no neutrino source terms are applied, and their angular momentum is set
to solid body rotation to eliminate viscous heating.  The inert matter in these
new cells is quickly swallowed by the BH, with a minimal impact on the
evolution.  This collapse procedure largely follows that of \cite{MF14} and
\cite{fahlman_2018}, allowing us to parameterize and isolate the HMNS lifetime
without needing to fine-tune many parameters in a microphysical EOS. 

For each simulation, we add $10^4$ passive, equal-mass tracer particles in the disk,
following the density distribution, in
order to record thermodynamic and kinematic quantities as a function of time.
In models with a finite HMNS lifetime, particles are added upon BH formation;
no disk material has left the domain by that time, so all relevant matter is sampled.
We designate trajectories associated to the unbound disk outflow as those that reach an
extraction radius $r=10^9$\,cm and have positive Bernoulli parameter
\begin{equation}
\label{eq:bernoulli}
Be = \frac{1}{2}\mathbf{v}^2 + e_{\rm int} + \frac{P}{\rho} + \Phi,
\end{equation}
with $\mathbf{v}$ the total fluid velocity, $e_{\rm int}$ the specific internal
energy, $P$ the total gas pressure, $\rho$ the mass density, and $\Phi$ the
gravitational potential.  These outflow trajectories are then post processed
with the nuclear reaction network code {\tt SkyNet} \cite{lippuner_skynet},
using the same settings as in \cite{lippuner_2017} and
\cite{Fernandez2020BHNS}.  The network employs $\sim 7800$ isotopes and more
than $10^5$ reactions, including strong forward reaction rates from the REACLIB
database \cite{cyburt_2010}, with inverse rates computed from detailed balance;
spontaneous and neutron-induced fission rates from \cite{frankel_1947},
\cite{mamdouh_2001}, \cite{wahl_2002}, and \cite{panov_2010}; weak rates from
\cite{fuller_1982}, \cite{oda_1994}, \cite{langanke_2000}, and the REACLIB
database; and nuclear masses from the REACLIB database, which includes
experimental values where available, or otherwise theoretical masses from the
finite-range droplet macroscopic model (FRDM) of \cite{moeller_2016}. 

%----------------------------------------------------------------------------------
\subsection{Neutrino Leakage Scheme and Flavor Transformation \label{s:leakage}}

We introduce a prescription to account for some of the salient features of
neutrino flavor transformation via the FFI in neutron star mergers. While we
directly solve neither the quantum kinetic equations nor the Boltzmann equation
for neutrinos, the following prescription is constructed to only transform
neutrino flavor outside of regions where neutrinos are in thermodynamic
equilibrium, since the angular asymmetries needed to incite the FFI are weak in
near-equilibrium conditions. We also provide a means to respect the
conservation of net lepton number called for by the symmetries of the neutrino
self-interaction potential.

\subsubsection{Leakage scheme}

The baseline neutrino leakage scheme used here follows \cite{ruffert_1996}, and
in particular the specific implementation described in
\cite{FM13,MF14,lippuner_2017}. While various modifications to the leakage
approach have been proposed to enhance its ability to realistically replicate
true neutrino transport (e.g., \cite{perego_2016,ardevol_2019}), our purpose
here is only to assess the potential impact of neutrino flavor transformation
in a variety of scenarios, and the computational efficiency of the present
scheme enables a large number of inexpensive simulations.  Nevertheless,
several upgrades have been made to the leakage implementation used in our
previous work (\cite{MF14}) in order to extend it to 3 species, borrowing from
the implementation in {\tt FLASH} reported in \cite{couch_2014}. First, a third
species (denoted by X) accounting for all heavy lepton species ($\nu_\mu,
\bar\nu_\mu, \nu_\tau, \bar\nu_\tau$) is now tracked. Second, emissivities due
to plasmon decay and electron-positron pair annihilation have been added for
all species, following \cite{ruffert_1996}.  Third, we compute opacities for
number and energy transport accounting for neutrino-nucleon elastic scattering
for all species, in addition to charged-current interactions for electron-type
neutrinos and antineutrinos, again following \cite{ruffert_1996}.  When
computing emissivities and opacities, the chemical potential for nucleons is
set to that of an ideal gas, for consistency with the equation of state used
(\S\ref{s:hydro}).  Finally, the electron neutrino and antineutrino chemical
potentials for Fermi blocking factors is obtained, as in \cite{ruffert_1996},
by interpolating between the beta equilibrium value for opaque regions and zero
for the transparent regime, but using the variable $\rho /
(10^{11}$\,g\,cm$^{-3})$ in lieu of optical depth. This is done to avoid an
iteration, since the optical depth depends on the opacity, which has Fermi
blocking factors.

The hydrodynamic source terms accounting for neutrino absorption are obtained
from the local absorption opacity and the incident luminosity of electron
neutrinos and antineutrinos. In our implementation, luminosities used for
absorption are made up of a contribution from the disk and another from the
HMNS, when present.  For disk luminosities, we use the annular lightbulb
prescription of \cite{FM13}, which heuristically accounts for neutrino
reabsorption by modeling incident radiation as originating from an equatorial ring with a
radius and luminosity representative of the net radiation produced by the disk.
In this prescription, the distribution function of emitted neutrinos is assumed
to have the form
\begin{equation}
\label{eq:distf_FM13}
f_{\nu_i} = e^{-\tau_{\mathrm{irr},i}}\frac{\mathcal{N}_{\nu_i}}{2\pi}\frac{\Theta(\cos\theta_k - \cos\theta_{k,{\rm min}})}{\exp(\epsilon/[kT_{\nu_i}])+1},
\end{equation} 
with
\begin{equation}
\label{eq:N_nu}
\mathcal{N}_{\nu_i} = \frac{L^*_{\nu_i}}{(7/16)4\pi r_{\rm em, i}^2 \sigma T_{\nu_i}^4}\,\,.
\end{equation}
Here, $\Theta$ is the step function, $\cos\theta_k$ is the angle between the
propagation direction and the radial direction, and $\cos\theta_{k,{\rm
min}}\simeq 1 - 0.5(r_{\rm em,i}/d)^2$.  The emission radius $r_{\rm em, i}$ is
an emissivity-weighted equatorial radius indicative of the point where most of
the neutrinos are emitted in the disk, while $d$ is the distance
between a point on this equatorial ring and the irradiated point. The neutrino
temperature $T_{\nu_i}$ is computed from the mean neutrino energy $\langle
\epsilon_{\nu_i}\rangle$, which in turn is obtained as in \cite{ruffert_1996}
by taking the ratio of the volume-integrated energy- to volume-integrated number emission rates
(we use the conversion $\langle \epsilon_{\nu_i}\rangle =
[\mathcal{F}_4(0)/\mathcal{F}_3(0)]\,kT_{\nu_i}\simeq 4\,kT_{\nu_i}$), with
$\mathcal{F}_i(\mu/kT)$ the Fermi-Dirac integral).  This prescription yields a
neutrino distribution that follows a Fermi-Dirac spectrum with temperature
$T_{\nu_i}$ and zero chemical
potential, but normalized such that the luminosity of the ring is equal to the
\emph{net} disk luminosity leaving the computational domain
$L^*_{\nu_i}=L_{\nu_i}-L_{\mathrm{abs},i}$. Here we denote the volume integral
of the neutrino emissivity $L_{\nu_i}$ and the volume integral of the neutrino
absorption power $L_{\mathrm{abs},i}$ \footnote{Absorption terms are computed
with the luminosity from the previous time step, and absorption terms are set
to zero during the first timestep after neutrino sources are turned on.}. In
previous work, it was sufficient to assume $L_{\mathrm{abs},i}=0$, since the
reabsorption correction produces no major qualitative changes 
on the dynamics and ejecta composition.
However, we find that accounting for $L_{\mathrm{abs},i}$ is needed to ensure
that the number luminosity of electron antineutrinos is higher than that of
electron neutrinos, as occurs for a leptonizing accretion disk, and the
relative number of different neutrino species does impact the effects of flavor
transformation.

The incident neutrino flux from the disk at any point $\mathbf{r}$ in the
computational domain is attenuated by a factor $\exp(-\tau_{\rm irr,i})$, where 
\begin{equation}
\label{eq:tau_irr}
\tau_{\rm irr,i}(\mathbf{r}) = \max[\tau_{\nu_i}(\mathbf{r}_{\rm em,i}), \tau_{\nu_i}(\mathbf{r})]
\end{equation} 
is the  maximum between the local optical depths at the emission maximum (annular ring 
$\mathbf{r}_{\rm em}$) and the irradiated point. The local optical depth at any location
is computed using the minimum between the vertical scale height, horizontal scale height,
and the radial direction
\begin{equation}
\label{eq:tau_local}
\tau_{\nu_i} = \kappa^{\rm e}_{\nu_i} \min(H_{\rm vert}, H_{\rm horiz}, r) 
\end{equation}
where $\kappa^{\rm e}_{\nu_i}$ is the neutrino opacity for energy transport,
$H_{\rm vert} = P/(\rho g |\cos\theta|)$ and $H_{\rm horiz} = P/(\rho
[g\sin\theta - a_{\rm cent}])$ are the vertical and horizontal scale heights,
respectively, with $g$ the
local acceleration of gravity, and $a_{\rm cent}$ the
centrifugal acceleration given the local specific angular momentum and
position.  See, e.g. \cite{ardevol_2019,fahlman_2022} for a comparison of this
optical depth prescription with others used in the literature.

The luminosity contribution from the HMNS, when present, is parametric and
imposed at the boundary. The following functional dependence is used (c.f.
\cite{MF14})
\begin{equation}
\label{eq:hmns_lum}
L^{\rm ns}_{\nu_e} = L^{\rm ns}_{\bar\nu_e} = L^{\rm ns}_{\nu_e,0}
                       \left[\frac{30\,\textrm{ms}}{\max(10\,\textrm{ms},t)}\right]^{1/2},
\end{equation}
with $L_{\nu_e,{\rm ns}}^0 = 2\times 10^{52}$\,erg\,s$^{-1}$. The normalization
of this functional form compares favorably with results obtained using moment
transport on the combined HMNS and disk system (e.g., Figure~3 of
\cite{fujibayashi_2020}), and the time dependence corresponds to diffusive
cooling \citep{salpeter_shapiro_1981}.  In our default setting, the heavy
lepton luminosity from the HMNS has the same time dependence and the same
normalization as the electron neutrinos and antineutrinos (i.e., $L^{\rm
ns}_{\rm X, 0} = L^{\rm ns}_{\nu_e,0}$).  To test the effect of this choice on
our results, for each HMNS lifetime, we run an additional model that increases
the heavy lepton luminosity normalization to twice the default value ($L^{\rm
ns}_{\rm X, 0} = 2L^{\rm ns}_{\nu_e,0}$). The neutrino temperatures of HMNS
neutrinos are constant and set to $T^{\rm ns}_{\nu_e} = 4$\,MeV, $T^{\rm
ns}_{\bar{\nu}_e} = 5$\,MeV, and $T^{\rm ns}_{\nu_x} = T^{\rm ns}_{\bar{\nu}_x}
= 6$\,MeV. This choice is made following typical values in protoneutron stars
(e.g., \cite{janka2001}). As with disk neutrinos, the spectrum is assumed to
follow a Fermi-Dirac distribution with zero chemical potential.

The local distribution function of neutrinos from the HMNS has a similar
functional form as equation (\ref{eq:distf_FM13}), with the following
differences \cite{MF14}: (1) there is no absorption correction to the
luminosity (i.e., $L_{\nu_i}^{*,\mathrm{ns}}=L_{\nu_i}^\mathrm{ns}$), (2)
the angular distribution is that of an emitting sphere, so we use the HMNS
radius instead of the ring radius and the factor $\cos\theta_{k,{\rm min}}$ is
computed analytically, (3) the neutrino temperatures are constant, and (4) the
attenuation factor uses the optical depth integrated along radial rays,
\begin{equation}
\label{eq:tau_ns}
\tau_{\nu_i}^{\rm ns}(\mathbf{r}) = \int_{r_{\rm ns}}^{r} \kappa_{\nu_i}^e(r,\theta) dr,
\end{equation}
with $r_{\rm ns}=30$\,km the stellar radius.  The neutrino absorption
contribution from the HMNS is then added to that from the disk. The energy
absorbed from HMNS neutrinos enters the absorption luminosity $L_{{\rm abs},i}$
used to correct the disk luminosity.

\subsubsection{Implementation of the FFI}

In the neutrino leakage treatment, emission and absorption of neutrinos are
treated separately. Flavor transformation occurs after emission during
propagation, so \emph{the neutrino emission terms are unchanged by flavor
transformation}. 

We include the effects of the FFI by modifying the incident neutrino fluxes and
neutrino temperatures for absorption. In order to restrict flavor
transformation to regions in the post-merger environment where we expect
instability (see, e.g., \cite{Just2022_FFI,richers_EvaluatingApproximateFlavor_2022}), we
control where flavor transformation occurs by interpolating between oscillated
and un-oscillated luminosities. At any point in the computational domain where
neutrino absorption takes place, the luminosity used in
equation~(\ref{eq:N_nu}) becomes
\begin{equation}
\label{eq:L_eff_heating}
L^*_{\nu_i}\to L^{\rm eff}_{\nu_i} = (1 - \eta_{\rm osc})L^*_{\nu_i} + \eta_{\rm osc}L^{\rm osc}_{\nu_i},
\end{equation}
where $L^*_{\nu_i}$ is the net un-oscillated luminosity, corrected for absorption,
and the superscript ``osc" indicates oscillated luminosities.
The activation parameter $\eta_{\rm osc}$ restricts flavor
transformation to regions where at least
one electron-type species is out of thermal equilibrium. Specifically, for disk luminosities we choose
\begin{equation}
\label{eq:eta_osc}
\eta_{\rm osc} = \exp({-\tau_{\bar{\nu}_e}}),
\end{equation}
where the local optical depth (equation~\ref{eq:tau_local}) to electron
antineutrinos is usually smaller than that to electron neutrinos, given the
lower proton fraction.

When a HMNS is present, the luminosities from the disk and the star are
oscillated separately, since in our formulation they originate from separate
locations. The oscillation parameter for the HMNS luminosities uses the same
radially-integrated optical depth used to attenuate it
(equation~\ref{eq:tau_ns}), i.e. $\eta_{\rm osc}^{\rm ns} = \exp(-\tau^{\rm
ns}_{\bar{\nu}_e})$.  This working definition results in a simple linear
superposition in regions transparent to both disk and HMNS neutrinos (polar
regions), while ignoring flavor transformation for HMNS neutrinos in regions
where they are heavily attenuated anyway (equator to mid-latitudes). In
\S\ref{s:results}, we show that disk luminosities are much larger than HMNS
luminosities and hence more impactful.

We express the flavor-transformed luminosities themselves as a linear
combination of the un-transformed luminosities:
\begin{eqnarray}
\label{eq:a_osc}
L^{\rm osc}_{\nu_e}       & = & (1 - a_{\rm osc})L^*_{\nu_e} + \,a_{\rm osc} L_{\nu_x}\\
\label{eq:b_osc}
L^{\rm osc}_{\bar{\nu}_e} & = & (1 - b_{\rm osc})L^*_{\bar{\nu}_e} + \,b_{\rm osc} L_{\bar{\nu}_x}.
\end{eqnarray}
We separate heavy lepton neutrinos from heavy lepton antineutrinos by evenly
splitting the total heavy lepton luminosity $L_{\rm X}$ produced by the leakage
scheme: $L_{\nu_x} = L_{\bar{\nu}_x} = (1/2) L_{\rm X}$.  This is justified in
that the mechanisms that produce heavy lepton neutrinos and antineutrinos are
symmetric. The electron neutrino and antineutrino temperatures for absorption
in equations (\ref{eq:distf_FM13})-(\ref{eq:N_nu}) are modified in the same way
as the luminosities
\begin{eqnarray}
\label{eq:Teff_nue}
kT^{\rm eff}_{\nu_e}       & = & \left(1 - \eta_{\rm osc}a_{\rm osc}\right)\,kT_{\nu_e}       + \eta_{\rm osc}a_{\rm osc}\,kT_{\nu_x}\\
\label{eq:Teff_nuebar}
kT^{\rm eff}_{\bar{\nu}_e} & = & \left(1 - \eta_{\rm osc}b_{\rm osc}\right)\,kT_{\bar{\nu}_e} + \eta_{\rm osc}b_{\rm osc}\,kT_{\bar{\nu}_x},
\end{eqnarray}
where $T_{\nu_x} = T_{\bar{\nu}_x} = T_{\rm X}$.
Reabsorption of heavy lepton neutrinos is neglected, since their absorption
opacities are much smaller than those of electron neutrinos and antineutrinos. 
Equations (\ref{eq:a_osc})-(\ref{eq:Teff_nuebar}) are applied separately to
disk and HMNS luminosities.

The coefficients $a_{\rm osc}$ and $b_{\rm osc}$ in equations
(\ref{eq:a_osc})-(\ref{eq:Teff_nuebar}) are scalar quantities that allow us to
manually tune how much flavor change occurs.  We test a variety of flavor
transformation assumptions:
\begin{enumerate}
\item \emph{Baseline:} $a_\mathrm{osc}=b_\mathrm{osc}=0$, which ensures no flavor transformation 
  and consistency with standard neutrino treatment.
                
\item \emph{Complete}: $a_\mathrm{osc}=b_\mathrm{osc}=1$, such that all neutrinos fully change flavor. 
  This is quite extreme and unrealistic.

\item \emph{Flavor Equilibration:} $a_\mathrm{osc}=b_\mathrm{osc}=2/3$ results in all neutrinos 
  and antineutrinos separately having equal abundances in all three flavors. This is still likely extreme.

\item \emph{Intermediate:} $a_\mathrm{osc}=b_\mathrm{osc}=1/2$ is a less extreme version of 
  the assumption of full Flavor Equilibration.

\item \emph{Asymmetric }(AS): The fast flavor instability is driven by the
  neutrino self-interaction Hamiltonian alone, the symmetries of which imply that the
  net lepton number cannot change. This requires that
  $a_\mathrm{osc}(N_{\nu_e}-N_{\nu_x}) =
  b_\mathrm{osc}(N_{\bar{\nu}_e}-N_{\bar{\nu}_x})$, with $N_{\nu_i}$ the local incident 
  number luminosity (Appendix~\ref{s:asymmetric}).
  We choose either  $a_\mathrm{osc}=2/3$ or $b_\mathrm{osc}=2/3$ and deem the other value
  \textit{asymmetric} as determined locally by this relationship. Given that
  electron neutrinos are generally sub-dominant by number, and therefore more likely
  to undergo flavor transformation, the case
  \begin{equation}
  \label{eq:b_AS}
  a_{\rm osc} = \frac{2}{3} \qquad b_{\rm osc} = \frac{2}{3}\left(\frac{N_{\nu_e} - N_{\nu_x}}{N_{\bar\nu_e}-N_{\bar{\nu}_x}}\right) 
  \end{equation}
  is expected to be the most realistic assumption. A related scheme was
  proposed in \cite{Just2022_FFI}.  In practice, we compute the asymmetric
  coefficient in equation~(\ref{eq:b_AS}) by using the global number luminosity
  attenuated with the appropriate optical depth (i.e., as in
  equation~\ref{eq:distf_FM13} for disk neutrinos), as geometric dilution cancels
  out. Also, the asymmetric coefficient is constrained to the range $[0,1]$.
\end{enumerate}

Note that Equations~(\ref{eq:a_osc})-(\ref{eq:b_osc}) allow both heavy
lepton neutrino flavors to transform to the electron-type flavor, instead of
restricting the flavor transformation to be between electron-type and
only one heavy lepton flavor. Our scheme conserves energy and neutrino number,
but does not reflect all of the symmetries of the Hamiltonian driving flavor
transformation. Because of this, in situations where there are equal numbers of
all three flavors (e.g., $L_X=2L_x = 2L_{\bar{x}}=4L_{\nu_e}=4L_{\bar{\nu}_e}$ when all flavors
have the same average energy), one would expect \emph{Complete} flavor transformation
($a_{\rm osc}=b_{\rm osc}=1$)
to leave all luminosities unchanged, since as many $\nu_e$ transform
into $\nu_\mu$ and vice versa, and likewise with $\nu_e-\nu_\tau$ as well as antineutrinos. 
In that situation, our scheme instead enhances the electron
neutrino luminosity to
$L_{\nu_e}^\mathrm{osc}=L_{\bar{\nu}_e}^\mathrm{osc}=2L_{\nu_e}^*$. NS
merger environments generally operate far from this limit, since there are
generally fewer heavy lepton neutrinos than electron neutrinos or antineutrinos, so
Equations~(\ref{eq:a_osc})-(\ref{eq:b_osc}) always result in a reduction of
electron flavor luminosity and we do not encounter this pathology. However, a
different construction (perhaps allowing only one heavy lepton flavor to
participate in transformation) may be needed to avoid pathologies in
environments like core-collapse supernovae, where heavy lepton neutrinos are
more abundant. While our HMNS luminosity prescription resembles the core-collapse
supernova regime, the disk luminosities
dominate throughout the evolution (c.f. Figure~\ref{fig:lum_ener_time}).

\begin{table*}
\caption{Models evolved and summary of results. Columns from left to right show
model name, oscillation coefficients $a_{\rm osc}$ and $b_{\rm osc}$ 
(equations~\ref{eq:a_osc}-\ref{eq:Teff_nuebar}), lifetime $t_{\rm ns}$ of the HMNS, 
ratio of HMNS heavy-lepton luminosity normalization to HMNS electron
neutrino luminosity normalization (equation~\ref{eq:hmns_lum}),
mass-outflow-averaged electron fraction and velocity at $r=10^9$\,cm
(equations~\ref{eq:ye_ave}-\ref{eq:vr_ave}), unbound mass ejected at
$r=10^9$\,cm, and unbound mass with $Y_e < 0.25$ that contributes to the red
kilonova component.
\label{t:models}} 
\begin{ruledtabular}
\begin{tabular}{lccccccccc}
Model & $a_{\rm osc}$        & $b_{\rm osc}$ & $t_{\rm ns}$ & $L^{ns}_{\rm X,0}$
      & $\langle Y_e\rangle$ & $\langle v_r\rangle$ & $M_{\rm ej}$ & $M_{\rm ej,red}$ \\
      &                      &               &     (ms)       & ($L^{ns}_{\nu_e,0}$)                    
      &                      & ($0.01$\,c)   &     ($10^{-2}\,M_\odot$)       & ($10^{-2}\,M_\odot$)   \\
\noalign{\smallskip}
\hline
BH-ab00   &  0  & 0   & 0         & --- & 0.29 & 2.8 & 2.4 & 0.08 \\
BH-ab05   & 1/2 & 1/2 &           &     & 0.28 & 3.1 & 1.9 & 0.09 \\
BH-ab07   & 2/3 & 2/3 &           &     & 0.27 & 3.2 & 1.9 & 0.18 \\
BH-ab10   & 1   & 1   &           &     & 0.26 & 3.4 & 1.8 & 1.09 \\
BH-aAS    & AS  & 2/3 &           &     & 0.27 & 3.1 & 2.0 & 0.29 \\
BH-bAS    & 2/3 & AS  &           &     & 0.28 & 3.1 & 1.8 & 0.15 \\
\noalign{\smallskip}
t010-ab00  & 0   & 0   & 10       & ---  & 0.27 & 3.0 & 3.2 & 0.69 \\
t010-ab05  & 1/2 & 1/2 &          &  1.0 & 0.26 & 3.2 & 2.9 & 0.54 \\
t010-ab07  & 2/3 & 2/3 &          &      & 0.26 & 3.2 & 2.4 & 1.29 \\
t010-ab10  & 1   & 1   &          &      & 0.24 & 3.1 & 2.3 & 1.72 \\
t010-aAS   & AS  & 2/3 &          &      & 0.25 & 2.9 & 2.6 & 1.16 \\
t010-bAS   & 2/3 & AS  &          &      & 0.26 & 3.1 & 2.5 & 0.62 \\
t010-L20   & 2/3 & AS  &          &  2.0 & 0.26 & 3.4 & 2.6 & 1.49 \\
\noalign{\smallskip}
t100-ab00  & 0   & 0   & 100      & ---  & 0.31 & 4.5 & 4.2 & 0.53 \\
t100-ab05  & 1/2 & 1/2 &          &  1.0 & 0.31 & 5.7 & 5.0 & 0.93 \\
t100-ab07  & 2/3 & 2/3 &          &      & 0.31 & 6.1 & 5.3 & 1.21 \\
t100-ab10  & 1   & 1   &          &      & 0.34 & 7.8 & 5.8 & 1.08 \\
t100-aAS   & AS  & 2/3 &          &      & 0.31 & 6.3 & 5.3 & 1.25 \\
t100-bAS   & 2/3 & AS  &          &      & 0.31 & 6.2 & 5.2 & 1.26 \\
t100-L20   & 2/3 & AS  &          &  2.0 & 0.31 & 6.1 & 5.1 & 1.23 \\
\noalign{\smallskip}
tinf-ab00  & 0   & 0   & $\infty$ & ---  & 0.38 & 7.3 & 9.7 & 0.51 \\
tinf-ab05  & 1/2 & 1/2 &          &  1.0 & 0.37 & 7.8 & 9.7 & 1.05 \\
tinf-ab07  & 2/3 & 2/3 &          &      & 0.38 & 8.2 & 9.6 & 1.10 \\
tinf-ab10  & 1   & 1   &          &      & 0.40 & 9.2 & 9.4 & 1.03 \\
tinf-aAS   & AS  & 2/3 &          &      & 0.38 & 8.2 & 9.6 & 1.31 \\
tinf-bAS   & 2/3 & AS  &          &      & 0.38 & 8.2 & 9.6 & 1.15 \\
tinf-L20   &     &     &          &  2.0 & 0.38 & 8.2 & 9.7 & 1.17 \\
tinf-noT\footnote{This model does not mix temperatures (eqns.~\ref{eq:Teff_nue}-\ref{eq:Teff_nuebar}).}  
             &     &     &          &  1.0 & 0.40 & 6.9 & 9.4 & 0.54 \\
\end{tabular}
\end{ruledtabular}
\end{table*}

%----------------------------------------------------------------------------------
\subsection{Models Evolved \label{s:models_evolved}}

All of our models are shown in Table~\ref{t:models}. We evolve four groups of
simulations that differ in the lifetime of the HMNS: $t_{\rm ns} = \{0, 10,100\}$\,ms, 
plus a set with a HMNS surviving until the end of the
simulation (labeled $t_{\rm ns}=\infty$).  All models are evolved for $17.735$\,s,
which corresponds to $5000$ orbits at $r=50$\,km (initial torus density maximum). By that
time, disks have lost at least 95\% of their initial mass to outflows and accretion.

For all four sets of models, we evolve neutrino flavor transformation cases
corresponding to \emph{Baseline}, \emph{Intermediate}, \emph{Complete},
\emph{Flavor Equilibration}, and \emph{Asymmetric}  (see \S\ref{s:leakage} for
definitions). Table~\ref{t:models} uses AS to refer to the coefficient set to
asymmetric, with the other held constant (e.g., $b_{\rm osc}= \textrm{AS}$
corresponds to equation~\ref{eq:b_AS}). The naming convention of models
indicates first whether it is a prompt BH or its HMNS lifetime, followed by the
value of the oscillation coefficients $a_{\rm osc}$ and $b_{\rm osc}$ if
symmetric (e.g., model t100-ab10 has $t_{\rm ns}=100$\,ms and $a_{\rm
osc}=b_{\rm osc}=1$) or by AS if one of them is set to asymmetric.  For each
HMNS lifetime, we also evolve models with $b_{\rm osc}=\textrm{AS}$ that double the
normalization of the HMNS heavy lepton luminosity (equation~\ref{eq:hmns_lum}), labeled
`L20'.  Additionally, we evolve a test model with $t_{\rm ns} = \infty$ and
$b_{\rm osc}=\textrm{AS}$ that includes transformation of neutrino fluxes
(Equations~\ref{eq:a_osc}-\ref{eq:b_osc}) but not neutrino mean energies
(Equations~\ref{eq:Teff_nue}-\ref{eq:Teff_nuebar}), denoted by tinf-noT.

%%%%%%%%%%%%%%%%%%%%%%%%%%%%%%%%%%%%%%%%%%%%%%%%%%%%%%%%%%%%%%%%%%%%%%%
\section{Results \label{s:results}}

\begin{figure*}
\centering
\includegraphics*[width=\linewidth]{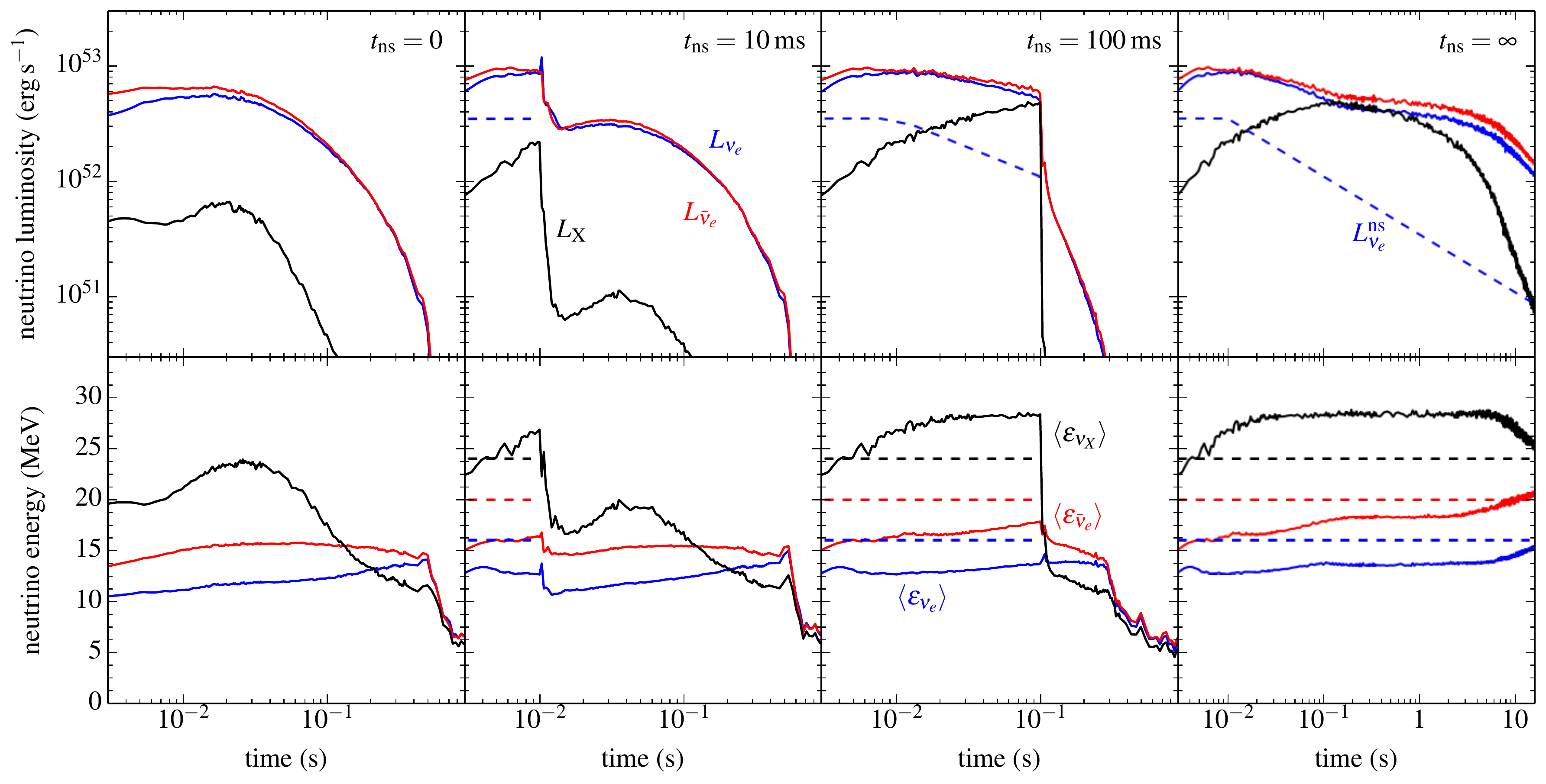}
\caption{Neutrino luminosities emitted by the disk (top) and associated mean
neutrino energies (bottom) in models without neutrino flavor transformation
($a_{\rm osc}=b_{\rm osc}=0$) and various HMNS lifetimes $t_{\rm ns}$, as
labeled (corresponding, from left to right, 
to models bh-ab00, t010-ab00, t100-ab00, and tinf-ab00 in Table~\ref{t:models}). 
The dashed lines show the imposed luminosity
(Equation~\ref{eq:hmns_lum}) and neutrino energies at the surface of the HMNS,
when present. Note that the disk luminosities used in Equation~(\ref{eq:N_nu})
are corrected for global absorption, and are thus lower than those shown here
(c.f. \S\ref{s:physical_origin}).
}
\label{fig:lum_ener_time}
\end{figure*}

\begin{figure*}
\centering
\includegraphics*[width=\linewidth]{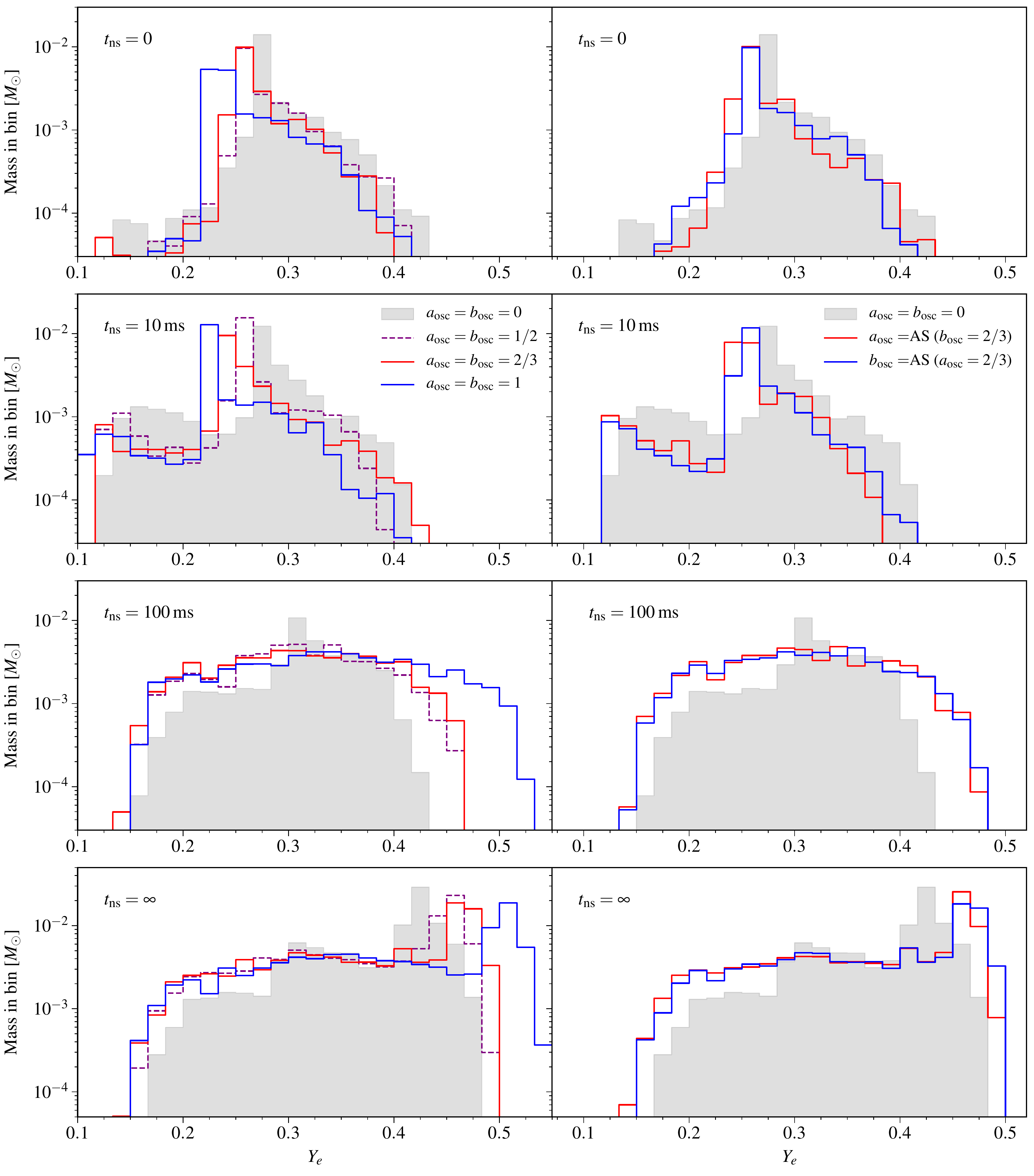} 
\caption{Mass histograms of electron fraction for unbound ejecta reaching
$r=10^9$\,cm by the end of each simulation ($t\simeq 17.7$\,s).  Rows from top
to bottom show groups of models with different HMNS lifetime $t_{\rm ns}$, as
labeled.  The gray shaded histograms show models without flavor transformation,
while colored curves show different combinations of flavor transformation
coefficients $\{a_{\rm osc},b_{\rm osc}\}$ (see \S\ref{s:leakage} for
definitions): runs with symmetric coefficients ($a_{\rm osc}=b_{\rm osc}$) are
on the left column, and asymmetric combinations on the right column. The bin
width is $\Delta Y_e \simeq 0.017$.}
\label{fig:ye_histogram}
\end{figure*}

\begin{figure*}
\centering
\includegraphics*[width=\linewidth]{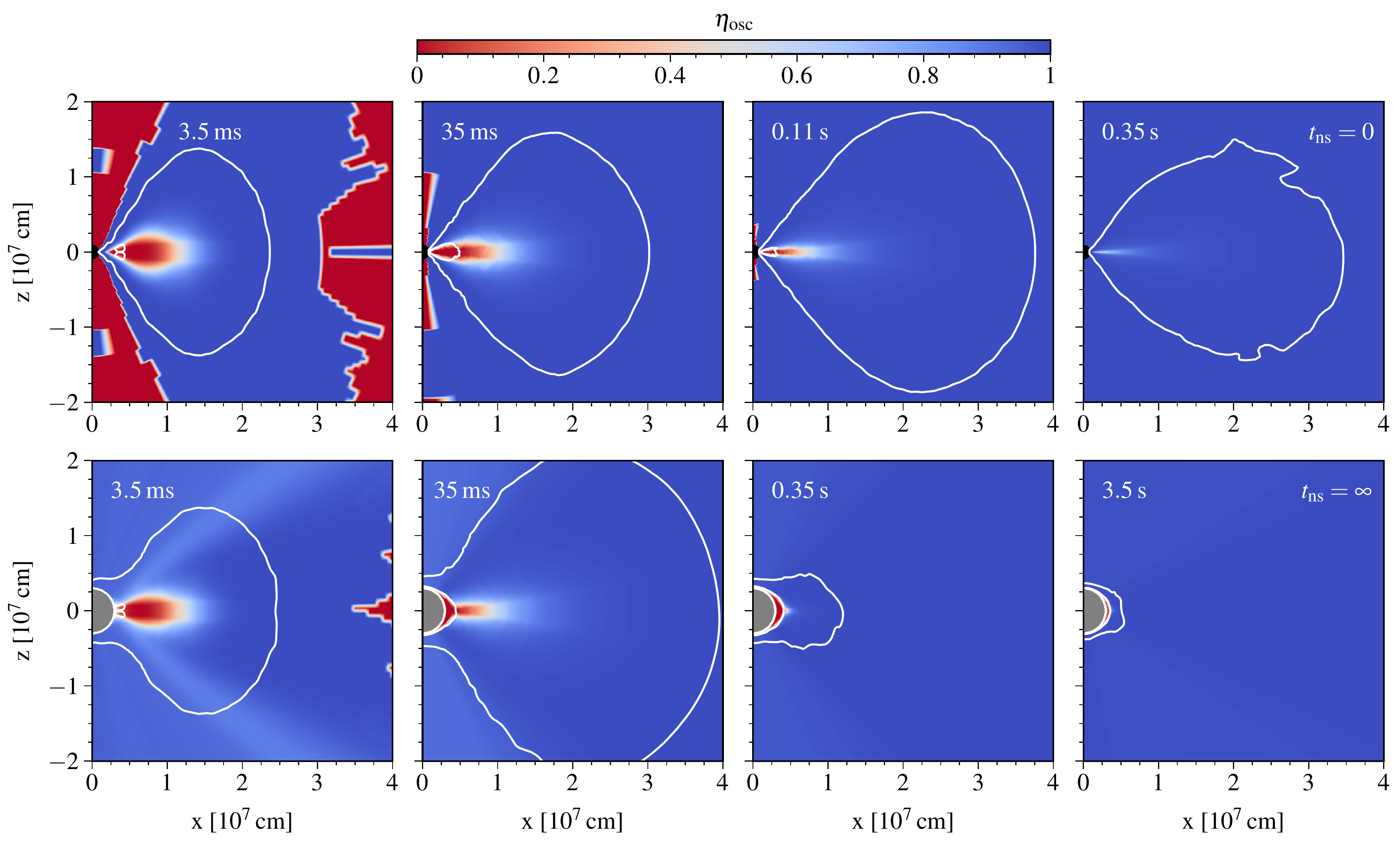}
\caption{Activation parameter $\eta_{\rm osc}$ (Equation~\ref{eq:eta_osc}) that
describes where we assume that the FFI operates, shown at selected times as
labeled in the upper left corner of each panel (the rotation
axis is along the $z$-axis and the equatorial plane is at $z=0$). The top row
shows the prompt BH model with no flavor transformation (BH-ab00), while the
bottom row shows the long-lived HMNS model with no flavor transformation
(tinf-ab00). For the latter, we compute -- in post-processing -- an effective
activation parameter that combines the contribution of disk and HMNS
luminosities (Equation~\ref{eq:eta_osc_eff}). The solid lines show density
contours at $10^{8}$\,g\,cm$^{-3}$ (outer) and $10^{11}$\,g\,cm$^{-3}$ (inner).
The black and gray circles indicate the size of the BH (absorbing) or HMNS
(reflecting) boundary. The square-edged red region in the leftmost panels, and
the cavity around the z-axis for the top row, corresponds to low-density
ambient material where neutrino source terms are suppressed and we set
$\eta_{\rm osc}=0$. }
\label{fig:eta-osc_snapshots}
\end{figure*}

%---------------------------------------------------------------
\subsection{Overview of Evolution without Flavor Transformation \label{s:overview}}

In order to analyze the effects of the FFI on the disk outflow, we first 
establish the baseline of comparison: accretion disks with variable HMNS lifetime 
that evolve without flavor transformation effects (model names ending in `ab00'). 
The initial maximum temperature and density in the torus
are $7\times 10^{10}$\,K ($\sim 6$\,MeV) and $8\times 10^{10}$\,g\,cm$^{-3}$,
respectively, thus neutrino emission from the disk is significant, and the disk
is optically thick in its densest regions (e.g., \cite{Ruffert+97,DiMatteo+02}).  

In the model with a promptly-formed BH (BH-ab00), the inner disk adjusts to a
near-Keplerian spatial distribution over a few orbits at $r=50$\,km (initial
density peak radius), with neutrino emission peaking at $t\sim 20$\,ms (top
left panel of Figure~\ref{fig:lum_ener_time}).  The emitted electron
antineutrino luminosity is slightly larger than the electron neutrino
luminosity, and both are about an order of magnitude larger than the combined
heavy lepton luminosity. Neutrino emission evolves on a timescale set by
viscous angular momentum transport, with luminosities dropping by a factor
$\sim 100$ below their maximum at a time $t\sim 400$\,ms.  Thereafter, the disk
is radiatively inefficient (e.g., \cite{Metzger+09a}).

When a HMNS is present, a boundary layer forms at the surface of the star, and
the disk can reach higher maximum densities and temperatures ($\sim
10^{12}$\,g\,cm$^{-3}$ and $\sim 10^{11}$\,K, respectively) than in the prompt
BH case. This results in electron neutrino and antineutrino luminosities from
the disk being higher by a factor of up to $\sim 2$ relative to the prompt BH
case.  For a long-lived HMNS (model tinf-ab00, top right panel of
Figure~\ref{fig:lum_ener_time}), disk luminosities decay much more slowly with
time than both the prompt BH luminosities and the HMNS luminosities imposed at
the boundary. The heavy lepton neutrino/antineutrino luminosity from the disk
$L_{\rm X}$ is significantly higher in model tinf-ab00 than in model BH-ab00,
rising to values within a factor of a few of the emitted electron neutrino and
antineutrino luminosities from the disk at $t\sim 200$\,ms.

The intermediate cases of a HMNS lasting for 10\,ms (model t010-ab00) or
100\,ms (model t100-ab00) show neutrino luminosities intermediate between the
prompt BH and long-lived HMNS cases.  In the model with $t_{\rm ns}=10$\,ms,
upon HMNS collapse, all luminosities drop sharply to a level below those of the prompt
BH case at the same time, and then recover over a timescale
$t\sim 100$\,ms until they approximately match those from model BH-ab00. 
The model with $t_{\rm ns}=100$\,ms is such that upon BH formation, all luminosities
also drop sharply but never recover to the level of the prompt BH model. We 
attribute this difference to transport of angular momentum by the boundary
layer when the HMNS is present. The chosen surface rotation period of $3$\,ms
corresponds to sub-Keplerian rotation at the stellar surface and
also at the ISCO radius of the BH, thus material co-rotating with the star at 
its surface is not able to circularize upon BH formation, and the resulting
disk has less matter at the same time than a torus that began evolving around a BH.

Before BH formation (and for $t \lesssim 100$\,ms in models BH-ab00 and
t010-ab00), the mean energy of heavy lepton neutrinos emitted by the disk is
higher by up to a factor of $\sim 2$ than those of electron neutrinos and
antineutrinos (bottom row of Figure~\ref{fig:lum_ener_time}).  This hierarchy
is due to the low opacity of heavy lepton neutrinos and the steeper temperature
dependence of the primary mechanism that produces them ($e^+e^-$ pair
annihilation). In all cases, the mean energies of electron antineutrinos
emitted by the disk are 20-50\% higher than the mean energies of electron
neutrinos, with values becoming close to one another only before a sharp drop
at $t\sim 0.5$\,s.  The drop in mean energies is a consequence of energy
luminosities decreasing faster with time than number luminosities as the disk
transitions to a radiatively inefficient state with lower temperature and
density. 

Due to enhanced neutrino irradiation and suppressed mass loss through the inner
boundary, a longer HMNS lifetime correlates with more mass ejected as well as
an overall higher average electron fraction and velocity of the unbound ejecta
\cite{MF14,Perego2014,fujibayashi2018,fahlman_2018,fujibayashi_2022}, which in
turn translates into a lower yield of heavy $r$-process elements
\cite{martin_2015,lippuner_2017,curtis_2021} (Table~\ref{t:models}).  Our model
t010-ab00 has a slightly lower average $Y_e$ than the prompt BH model due to a
relative increase in the ejecta with  $Y_e < 0.25$ material
(Figure~\ref{fig:ye_histogram})

Mass ejection in our models is driven by neutrino energy deposition, viscous
heating, and nuclear recombination. Neutrino-driven outflows operate on a
timescale of $\gtrsim 10$\,ms and are significant whenever a HMNS is present.
In pure BH models, and also in late-time HMNS disks, mass is primarily ejected
by a combination of viscous heating and nuclear recombination, operating on a
timescale of few $100$\,ms. Simulations that include MHD effects have
additional mass ejection channels available in the form of magnetic stresses
(Lorentz force) that eject matter on a $\sim$\,ms timescale, providing a
distinct component (e.g.,
\cite{siegel_2018,F19_grmhd,miller2019,Just2022_Yeq,hayashi_2022}).  The
composition of this prompt disk outflow is sensitive to that of the disk upon
formation (i.e., neutron-rich), since weak interactions do not operate for long
enough to bring $Y_e$ toward its equilibrium value. The properties of this
early magnetic-driven ejection component are also sensitive to the initial
field geometry (e.g., \cite{christie_2019,fahlman_2022}).  In models with a
long-lived HMNS, magnetic stresses and/or neutrino absorption combine to launch
a fast outflow (e.g., \cite{Siegel_2014,ciolfi_2019,moesta_2020,shibata_2021,most_2021,sun_2022}). 

\begin{figure*}
\includegraphics*[width=\textwidth]{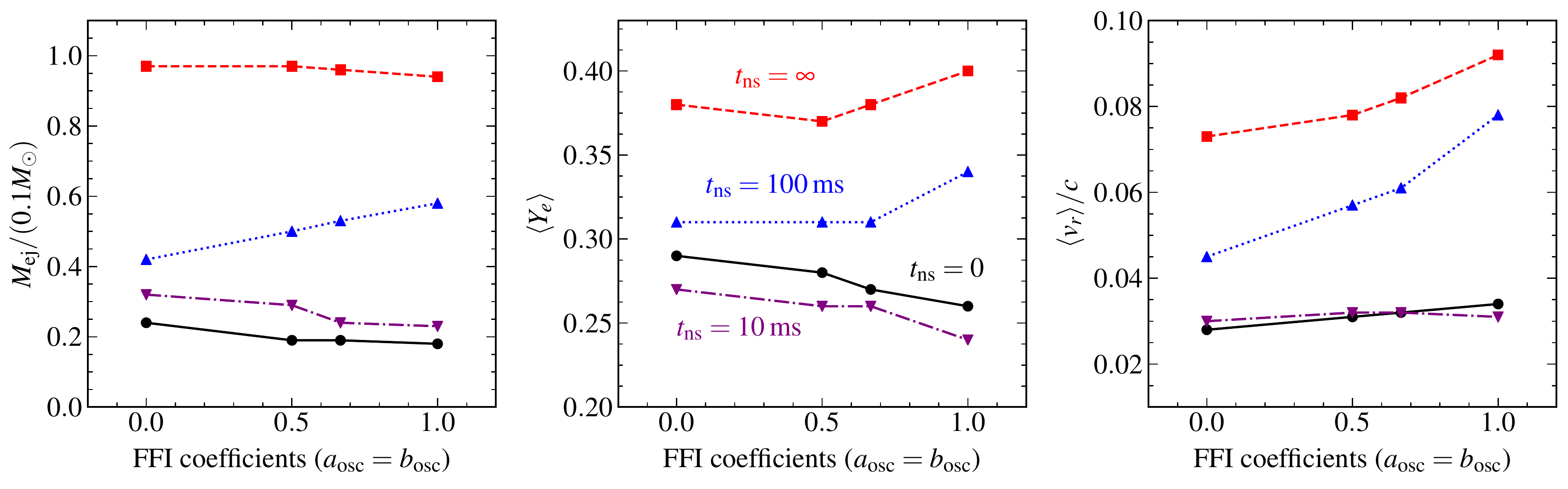}
\caption{Average outflow properties as a function of flavor transformation
intensity, parameterized through the FFI coefficients $\{a_{\rm osc},b_{\rm
osc}\}$ in symmetric combinations ($a_{\rm osc}=b_{\rm osc}$,
Table~\ref{t:models}).  Shown are the total unbound ejected mass (left),
average electron fraction (middle), and average radial velocity (right), for
various HMNS lifetimes $t_{\rm ns}$, as labeled in the middle panel.  }
\label{fig:table_results}
\end{figure*}

%-----------------------------------------------------------------------------
\subsection{Effect of Flavor Transformation on Outflow Properties \label{s:flavor_transformation}}

\subsubsection{Overall Trends}

Figure~\ref{fig:eta-osc_snapshots} shows the FFI activation parameter
$\eta_{\rm osc}$ (Equation~\ref{eq:eta_osc}) for the prompt BH and long-lived
HMNS models with $a_{\rm osc} = b_{\rm osc} = 0$ at various times in the
evolution.  The BH disk starts optically thick in its denser regions and flavor
transformation operates outside these opaque regions, by construction. For as
long as neutrino emission remains significant, the disk retains a dense core
where flavor transformation does not operate, while $\eta_{\rm osc}\sim 1$ in
all the outflow material. 

To diagnose the long-lived HMNS case, we compute an effective activation parameter
(weighted by attenuated luminosity) that combines disk 
and stellar contributions (which are treated separately
in our formalism, see \S\ref{s:leakage}):
\begin{equation}
\label{eq:eta_osc_eff}
\eta_{\rm osc}^{\rm (eff)} \equiv \frac{L^*_{\bar{\nu}_e}\,\eta_{\rm osc}^2 + L^{\rm ns}_{\bar{\nu}_e}\,{\eta^{\rm ns}_{\rm osc}}^2 }
                                      {L^*_{\bar{\nu}_e}\,\eta_{\rm osc} + L^{\rm ns}_{\bar{\nu}_e}\,\eta^{\rm ns}_{\rm osc} }.
\end{equation}
This formulation implicitly neglects the difference between the distance to the
HMNS surface and the disk emission ring, and assumes $\tau_{{\rm
irr},\bar{\nu}_e} = \tau_{\bar{\nu}_e}$ (Equation~\ref{eq:tau_local}).  The
disk optical depth is initially the same as in the BH case, but as accretion
proceeds,  a dense and neutrino-opaque boundary layer forms at the surface of
the HMNS.  Figure~\ref{fig:eta-osc_snapshots} shows that most of the disk and
its outflow have nevertheless $\eta_{\rm osc}\sim 1$, which is due to the
dominance of disk luminosities over HMNS luminosities (c.f.,
Figure~\ref{fig:lum_ener_time}). In fact, the opaque boundary layer prevents
neutrinos emitted from the HMNS surface from reaching the disk, from the
equator up to mid-latitude regions. Neutrino emission from the disk, on the
other hand, is optically thin everywhere except the disk midplane at early
times and the boundary layer, whenever present. This suggests that the effects of
the FFI manifest primarily through disk luminosities on equatorial 
latitudes, while a mixture of both contributions acts along polar latitudes.

To quantitatively assess the effects of flavor transformation on our model set,
we describe global outflow properties through quantities measured at an
extraction radius $r_{\rm out}=10^9$\,cm. The total ejected mass with positive
Bernoulli parameter (Equation~\ref{eq:bernoulli}) reaching that radius over the
course of the simulation is denoted by $M_{\rm ej}$, and the subset of this
mass with $Y_e < 0.25$ by $M_{\rm ej,red}$. The average electron fraction and
radial velocity at $r_{\rm out}$ are weighted by the mass-flux (e.g.,
\cite{FM13})
\begin{eqnarray}
\label{eq:ye_ave}
\langle Y_e \rangle & = & \frac{\int r_{\rm out}^2 \rho v_r Y_e\, d\Omega\,dt}{\int r_{\rm out}^2 \rho v_r\, d\Omega\, dt}\\
\label{eq:vr_ave}
\langle v_r \rangle & = & \frac{\int r_{\rm out}^2 \rho v_r^2\, d\Omega\,dt}{\int r_{\rm out}^2 \rho v_r\, d\Omega\, dt},
\end{eqnarray}
where only matter with positive Bernoulli parameter is included in the
integral, and the time range includes the entire simulation.

Table~\ref{t:models} shows that for each HMNS lifetime, the overall changes
introduced by neutrino flavor transformation on the ejecta properties are
moderate: at most $\sim 10\%$ in average electron fraction, up
to $\sim 40\%$ in total ejecta mass, and $\sim 10-40\%$ in average velocity
except for the most extreme FFI case with $t_{\rm ns}=100$\,ms, for which it
is a $73\%$ increase. The mass with $Y_e < 0.25$ ($M_{\rm ej,red}$) can
change by a factor of up to a few.

The direction of these changes depends on the HMNS lifetime, as illustrated by
Figure~\ref{fig:table_results} for models with symmetric FFI coefficients
($a_{\rm osc}=b_{\rm osc}$). For $t_{\rm ns}\leq 10$\,ms, the average electron
fraction of models with flavor transformation is always lower than in the
un-oscillated case. Figure~\ref{fig:ye_histogram} illustrates the shift of the
electron fraction distribution to more neutron-rich values, with the peak of
mass ejection typically decreasing by up to $0.05$.
A HMNS with lifetime $t_{\rm ns}\geq 100$\,ms, on the other hand, shows an
overall broadening of the $Y_e$ distribution when including flavor
transformation, with the average electron fraction staying constant
or increasing by at most $0.02$. The long-lived HMNS set shows
a peak $Y_e$ value shifting to higher, proton-rich values.

In all cases, the average outflow velocity stays nearly constant or increases
(most notably for $t_{\rm ns}=100$\,ms) when including flavor transformation
relative to the baseline case. Likewise, mass ejection decreases with 
a more intense FFI all in cases except for the set with $t_{\rm ns}=100$\,ms.

For symmetric values of the oscillation coefficients (model names ending in
ab00, ab05, ab07, and ab10), the magnitude of the changes introduced by flavor
transformation generally varies monotonically with the value of these
coefficients.  The case $a_{\rm osc} = b_{\rm osc}=1$ shows the strongest
effect, as expected.  When using asymmetric coefficients, we find that the
ratio of number luminosities in equation~(\ref{eq:b_AS}) starts low, since
initially $N_{\nu_e}<N_{\bar{\nu_e}}$, but approaches values close to unity on a
timescale of $\sim 35$\,ms (10 orbits at $r=50$\,km).  The magnitude of the
changes in average quantities is similar (but not always identical) to the case
$a_{\rm osc} = b_{\rm osc} = 2/3$, as expected.  Differences between setting
either $a_{\rm osc}$ or $b_{\rm osc}$ to asymmetric are minor, as shown by the
right column of Figure~\ref{fig:ye_histogram}.

\begin{table*}
\caption{Average of quantities extracted from tracer particles.  From left to
right, the first 9 columns show model name, net change in $Y_e$
(eq.~\ref{eq:dye_net}) , change in $Y_e$ due to emission/absorption of electron
neutrinos/antineutrinos (eqns.~\ref{eq:dye_definition}-\ref{eq:ghep}, as
labeled), net energy change due to viscous heating, neutrino
emission/absorption, and alpha particle recombination
(eqns.~\ref{eq:dq_visc}-\ref{eq:dq_alpha}). 
For models with $t_{\rm ns}=10$\,ms and $100$\,ms, direct time integration of
source terms (columns 2-9) is done from BH formation onward ($t_{\rm
min}=t_{\rm ns})$.  Likewise, for the long-lived HMNS model, integration begins
at $t_{\rm min}=0$ but stops at $t_{\rm max}=35$\,ms.  Therefore, except for
the prompt BH series, the net change in $Y_e$ in the second column does not
capture the entire history, and comparisons should only be made between
simulations with the same $t_{\rm ns}$. 
The last two columns show the
average mass fraction of Lanthanides ($X_{\rm La}$) and Actinides ($X_{\rm
Ac}$) obtained with {\tt SkyNet} including the entire simulation, and
extrapolating the trajectories to $30$\,yr. 
\label{t:analysis}} 
\begin{ruledtabular}
\begin{tabular}{lcccccccccc}
Model & $\Delta \bar{Y}^{\rm net}_e$ & $\Delta \bar{Y}^{{\rm em},\nu_e}_e$ & $\Delta \bar{Y}^{{\rm em},\bar{\nu}_e}_e$ &
        $\Delta \bar{Y}^{{\rm abs},\nu_e}_e$ & $\Delta \bar{Y}^{{\rm abs},\bar{\nu}_e}_e$ &
        $\Delta \bar{q}_{\rm visc}$ & $\Delta \bar{q}_{\nu}$ & $\Delta \bar{q}_{\alpha}$ & 
        $\bar{X}_{\rm La}$ & $\bar{X}_{\rm Ac}$\\
      & & & & & & \multicolumn{3}{c}{($10^{19}$\,erg\,g$^{-1}$)} & $(10^{-2})$ & $(10^{-2})$\\
\noalign{\smallskip}
\hline
BH-ab00   & 0.19 & 1.05 & 0.96 & 0.44 & 0.16 & 2.06 & -1.07 & 0.33 & 0.8 & 0.3\\ 
BH-ab05   & 0.18 & 1.03 & 0.99 & 0.34 & 0.12 & 2.13 & -1.15 & 0.32 & 0.8 & 0.2\\
BH-ab07   & 0.17 & 1.08 & 1.06 & 0.30 & 0.11 & 2.31 & -1.31 & 0.31 & 0.9 & 0.2\\
BH-ab10   & 0.16 & 1.09 & 1.12 & 0.20 & 0.08 & 2.60 & -1.58 & 0.30 & 1.9 & 0.3\\
BH-aAS    & 0.17 & 1.04 & 1.06 & 0.25 & 0.11 & 2.43 & -1.34 & 0.31 & 1.0 & 0.3\\
BH-bAS    & 0.17 & 1.08 & 1.07 & 0.30 & 0.13 & 2.36 & -1.30 & 0.31 & 1.1 & 0.2\\
\noalign{\smallskip}
t010-ab00\footnote{For this group of models, source terms are integrated over $t\geq t_{\rm ns}=10$\,ms.}  
          & 0.11 & 0.69 & 0.58 & 0.32 & 0.10 & 1.39 & -0.56 & 0.29 & 2.3 & 0.9\\
t010-ab05  & 0.10 & 0.63 & 0.57 & 0.22 & 0.07 & 1.50 & -0.65 & 0.28 & 1.5 & 1.3\\
t010-ab07  & 0.09 & 0.59 & 0.56 & 0.17 & 0.05 & 1.52 & -0.67 & 0.28 & 2.1 & 1.4\\
t010-ab10  & 0.07 & 0.57 & 0.58 & 0.08 & 0.03 & 1.67 & -0.83 & 0.26 & 3.0 & 1.1\\
t010-aAS   & 0.09 & 0.54 & 0.54 & 0.14 & 0.05 & 1.44 & -0.66 & 0.27 & 2.1 & 1.4\\
t010-bAS   & 0.09 & 0.61 & 0.59 & 0.18 & 0.06 & 1.63 & -0.70 & 0.28 & 1.8 & 1.3\\
t010-L20   & 0.09 & 0.65 & 0.62 & 0.19 & 0.06 & 1.63 & -0.76 & 0.28 & 1.8 & 1.4\\
\noalign{\smallskip}
t100-ab00\footnote{For this group of models, source terms are integrated over $t\geq t_{\rm ns}=100$\,ms.}
           & 0.011 & 0.029 & 0.035 & 0.009 & 0.004 & 0.25 & -0.04 & 0.13 & 0.6 & 0.08 \\
t100-ab05  & 0.004 & 0.009 & 0.013 & 0.001 & 7E-4  & 0.14 & -0.01 & 0.10 & 1.1 & 0.2 \\
t100-ab07  & 0.002 & 0.006 & 0.008 & 6E-4  & 3E-4  & 0.09 & -0.01 & 0.09 & 1.1 & 0.3 \\
t100-ab10  & 0.001 & 0.004 & 0.005 & 3E-5  & 1E-5  & 0.06 & -8E-3 & 0.07 & 1.1 & 0.2 \\
t100-aAS   & 0.003 & 0.009 & 0.011 & 8E-4  & 4E-4  & 0.09 & -0.02 & 0.09 & 1.2 & 0.3 \\
t100-bAS   & 0.003 & 0.007 & 0.010 & 6E-4  & 4E-4  & 0.12 & -0.01 & 0.09 & 1.2 & 0.3 \\
t100-L20   & 0.003 & 0.010 & 0.013 & 9E-4  & 5E-4  & 0.11 & -0.02 & 0.09 & 1.3 & 0.3 \\
\noalign{\smallskip}
tinf-ab00\footnote{For this group of models, source terms are integrated over $0 \leq t\leq 35$\,ms.} 
          & 0.08 & 1.00 & 0.88 & 0.41 & 0.20 & 1.13 & -1.79 & 0.02 & 0.5 & 0.2\\
tinf-ab05 & 0.08 & 1.02 & 0.90 & 0.40 & 0.19 & 1.09 & -1.63 & 0.02 & 1.0 & 0.4\\
tinf-ab07 & 0.09 & 1.02 & 0.90 & 0.40 & 0.19 & 1.08 & -1.52 & 0.02 & 0.9 & 0.4\\
tinf-ab10 & 0.11 & 1.06 & 0.90 & 0.43 & 0.17 & 1.07 & -1.40 & 0.02 & 0.7 & 0.2\\
tinf-aAS  & 0.09 & 1.04 & 0.92 & 0.41 & 0.19 & 1.08 & -1.47 & 0.02 & 1.0 & 0.4\\
tinf-bAS  & 0.09 & 1.01 & 0.90 & 0.40 & 0.19 & 1.09 & -1.52 & 0.02 & 0.9 & 0.4\\
tinf-L20  & 0.09 & 1.05 & 0.91 & 0.44 & 0.21 & 1.09 & -1.49 & 0.02 & 0.9 & 0.4\\
tinf-noT  & 0.07 & 1.00 & 0.87 & 0.37 & 0.17 & 1.15 & -1.91 & 0.01 & 0.4 & 0.1 
\end{tabular}
\end{ruledtabular}

\end{table*}

\subsubsection{Physical Origin of the Changes due to the FFI \label{s:physical_origin}}

The effect of the FFI on the disk outflow can be ultimately traced back to the
hierarchy of luminosities and energies shown in Figure~\ref{fig:lum_ener_time}.
For the prompt BH case, where only neutrinos from the disk exist, flavor
transformation through equations (\ref{eq:a_osc}), (\ref{eq:b_osc}),
(\ref{eq:Teff_nue}), and (\ref{eq:Teff_nuebar}) replaces a high-luminosity,
low-energy species ($\nu_e$, $\bar{\nu}_e$) for a low-luminosity, high-energy
species ($\nu_x$, $\bar{\nu}_x$). In the optically-thin limit, neutrino number
absorption is proportional to $\sim L_{\nu_i} \langle \epsilon_{\nu_i}\rangle$,
with energy absorption having an an additional power\footnote{For simplicity, 
we assume $\langle \epsilon_{\nu_i}^2\rangle=\langle \epsilon_{\nu_i}\rangle^2$ 
in the argument.} of $\langle \epsilon_{\nu_i}\rangle$.
In this simple picture, a complete flavor swap should decrease the
electron-flavor neutrino luminosity by an order of magnitude and increase the
average energy of electron-flavor neutrinos by a factor of up to $\sim 2$, for
an overall decrease in number absorption of a factor of $\sim 2$.

\begin{figure*}
\includegraphics*[width=\linewidth]{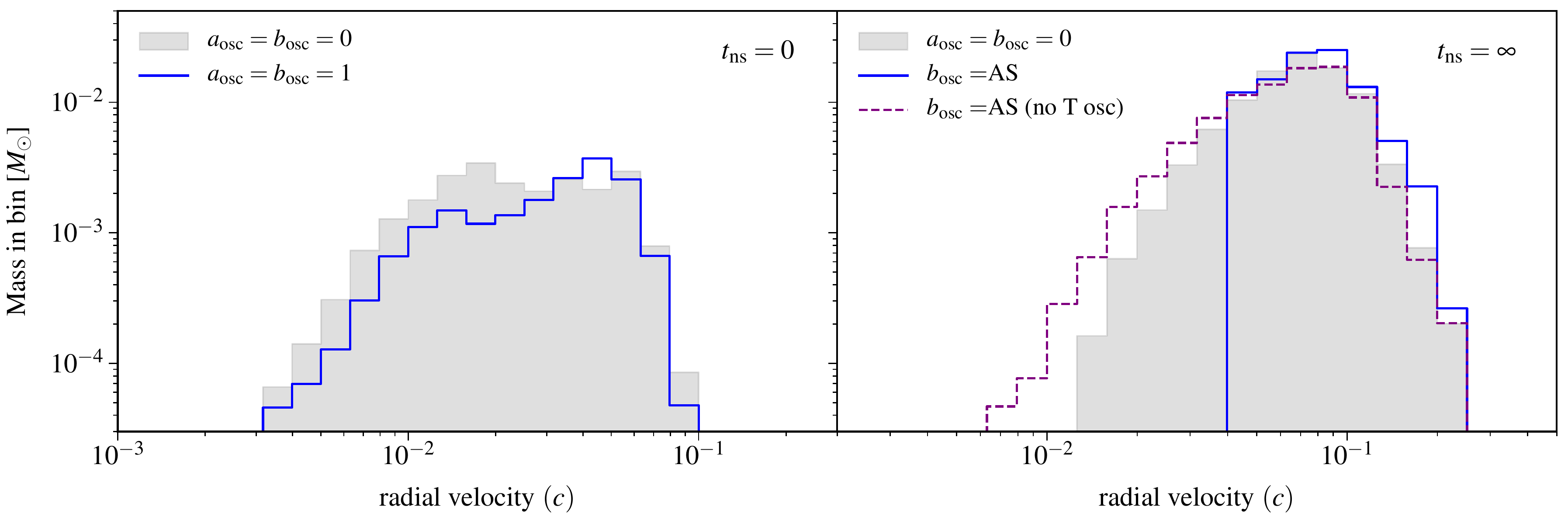}
\caption{Mass histograms of radial velocity for unbound ejecta from selected
simulations with a prompt BH (left) and long-lived HMNS (right), with flavor
transformation coefficients as labeled. The dashed line on the right
panel corresponds to the long-lived HMNS model with asymmetric
$b_{\rm osc}$ but no mixing of the neutrino temperatures 
(model tinf-noT in Tables~\ref{t:models} and \ref{t:analysis}).
The bin width is $\Delta \log (v_r/c) = 0.1$.}
\label{fig:hist_vel}
\end{figure*}

To diagnose quantitatively the effects on the electron fraction, we show in
Table~\ref{t:analysis} the time-integral of the source terms that control the
evolution of $Y_e$ (c.f., \cite{Fernandez2020BHNS}),
\begin{equation}
\label{eq:dye_definition}
\Delta Y_e^i = \int_{t_{\rm min}}^{t_{\rm max}} \Gamma^i dt,
\end{equation}
where $\Gamma^i$ is the rate per baryon of charged-current weak processes:
\begin{eqnarray}
\label{eq:gcem}
\Gamma^{\rm em, \nu_e} = \lambda_{e^-}Y_p:        && e^- + p \to n + \nu_e\\
\label{eq:gcep}
\Gamma^{\rm em, \bar{\nu}_e} = \lambda_{e^+}Y_n:  && e^+ + n \to p + \bar{\nu}_e\\
\label{eq:ghem}
\Gamma^{\rm abs, \nu_e}=\lambda_{\nu_e}Y_n:       && \nu_e + n \to  e^- + p\\
\label{eq:ghep}
\Gamma^{\rm abs, \bar{\nu}_e} = \lambda_{\bar{\nu}_e}Y_p: && \bar{\nu}_e + p \to  e^+ + n
\end{eqnarray}
where $\lambda_i$ are the reaction rates per target particle, and $Y_{n,p}$ the
number of neutrons or protons per baryon (in the notation of
\cite{Just2022_FFI}).  Equation~(\ref{eq:dye_definition}) is computed for each
weak process in each trajectory that is unbound and reaches $r>10^9$\,cm.
Values are then averaged arithmetically over trajectories (which have identical
mass for a given run), denoted by a bar above, and the net change is computed 
\begin{equation}
\label{eq:dye_net}
\Delta \bar{Y}^{\rm net}_e = \Delta \bar{Y}^{\rm em, \bar{\nu}_e}_e - \Delta \bar{Y}^{\rm em, \nu_e}_e 
                             + \Delta \bar{Y}^{\rm abs, \nu_e}_e - \Delta \bar{Y}^{\rm abs, \bar{\nu}_e}_e.
\end{equation}
The time range $[t_{\rm min},t_{\rm max}]$ in
equation~(\ref{eq:dye_definition}) is different for model sets with different
HMNS lifetime. For the prompt BH case, the interval is the entire simulation
time. For sets with $t_{\rm ns} = 10$\,ms and $100$\,ms, the interval begins at
BH formation ($t_{\rm min}=t_{\rm ns}$) and extends to the end of the
simulation, because particles are created after HMNS collapse. For these model
sets, quantities capture the impact of the FFI on post-collapse neutrino
processes. For the long-lived HMNS case, the period extends from the start of
the simulation ($t_{\rm min}=0$) until $10$ orbits at $r=50$\,km ($t_{\rm
max}\simeq 35$\,ms).  Limiting the integration interval is necessary because
the absorption and emission terms in the long-lived HMNS case are large, and
our trajectories are sampled at coarser time intervals at later times, leading
to imprecise cancellation of large terms when numerically integrating over very
long time intervals in post-processing. Because of this, direct comparisons
should be made across models with the same $t_\mathrm{ns}$ in Table~\ref{t:analysis}. Note that the
tracer particles themselves are updated every time step and do not suffer from
this post-processing error. In all cases, the chosen time interval satisfies
$\bar{Y}_e(t_{\rm min})+\Delta \bar{Y}^{\rm net}_e = \bar{Y}_e (t_{\rm  max})$.

The change in $Y_e$ with flavor transformation in the prompt BH models can be
explained straightforwardly: as the FFI becomes stronger for
models with increasing values of $a_{\rm osc}=b_{\rm osc}$, the average
absorption of both electron neutrinos and antineutrinos decreases, with $\Delta
\bar{Y}^{\rm abs, \nu_e}_e$ decreasing more than $ \Delta \bar{Y}^{\rm abs,
\bar{\nu}_e}_e$, with a maximum drop in absorption of a factor $\sim 2$.
Emission terms, on the other hand, change by a few percent at most, since
flavor transformation only changes emission rates indirectly through a minor
effect on the disk dynamics. A decrease in neutrino absorption decreases the
rate at which weak interactions bring $Y_e$ to its equilibrium value, and also
lowers the equilibrium value itself (e.g., \cite{Just2022_Yeq,Just2022_FFI}).

We diagnose the effects of flavor transformation on the outflow dynamics by
integrating energy source terms of fluid elements.  Table~\ref{t:analysis}
shows the average specific energy gain of disk outflow trajectories through the
quantities
\begin{eqnarray}
\label{eq:dq_visc}
\Delta q_{\rm visc} & = & \int_{t_{\rm min}}^{t_{\rm max}} \dot{q}_{\rm visc} dt\\
\label{eq:dq_nu}
\Delta q_\nu        & = & \int_{t_{\rm min}}^{t_{\rm max}} \dot{q}_{\rm \nu} dt\\
\label{eq:dq_alpha}
\Delta q_\alpha     & = & \frac{B_\alpha}{m_\alpha}\left [X_{\alpha}(t_{\rm max}) - X_\alpha(t_{\rm min}) \right],
\end{eqnarray}
where $ \dot{q}_{\rm visc}$ and $\dot{q}_{\rm \nu}$ are the rate of viscous
and net neutrino heating per unit mass, respectively, ${B_\alpha}/{m_\alpha}$ is
the nuclear binding energy per unit mass of alpha particles, and $X_{\alpha}$
is the mass fraction of alpha particles. For each model, the time range and
particle sample employed is the same as in equation~(\ref{eq:dye_definition}).

For the prompt BH models, the overall decrease in the neutrino absorption to
emission ratio due to flavor transformation also results in higher net cooling
of the torus (more negative $\Delta \bar{q}_\nu$), decreasing the vertical
extent of the disk.  This is accompanied by an increase in viscous heating,
which is proportional to the local disk pressure (c.f., \cite{FM13}).  In the
BH model with \emph{Complete} flavor transformation (BH-ab10), net neutrino
cooling increases by $60\%$ while viscous heating increases by $30\%$ relative
to the \emph{Baseline} model (BH-ab00). The left panel of
Figure~\ref{fig:hist_vel} shows the velocity distribution of the outflow from
both models: the high-velocity portion of the histogram remains at a similar
level, with flavor transformation inducing a slight shift to higher velocities
of the peak, and a sizable decrease in the low velocity portion. If we change
the unbinding criterion from Bernoulli to positive escape velocity\footnote{The
positive escape velocity criterion is more stringent, showing less overall mass
ejected, and selecting only higher-velocity matter. The Bernoulli criterion
accounts for the conversion of internal energy to bulk kinetic energy via
adiabatic expansion, allowing slower matter to be considered as having
sufficient energy to become gravitationally unbound.}, we find that the amount
of mass ejected in model BH-ab10 is $10\%$ \emph{higher} than in model BH-ab00,
versus $25\%$ lower if we use the default Bernoulli criterion. Thus, the
overall change in energy source terms introduced by the FFI in BH disks results
in \emph{less marginally unbound mass ejected}.

Regarding the long-lived HMNS model set (tinf), Table~\ref{t:analysis} shows
that absorption of electron neutrino number slightly \emph{increases} or stays
constant in models with increasing $a_{\rm osc}=b_{\rm osc}$ in the first
$35$\,ms of evolution, while absorption of electron antineutrino number
decreases in a similar way as in the pure BH case.  A decrease in electron
antineutrino absorption relative to neutrino absorption, with little change in
the emission terms, increases the equilibrium value toward which $Y_e$ is
driven (e.g., \cite{Just2022_Yeq}).

\begin{figure}
\includegraphics*[width=\columnwidth]{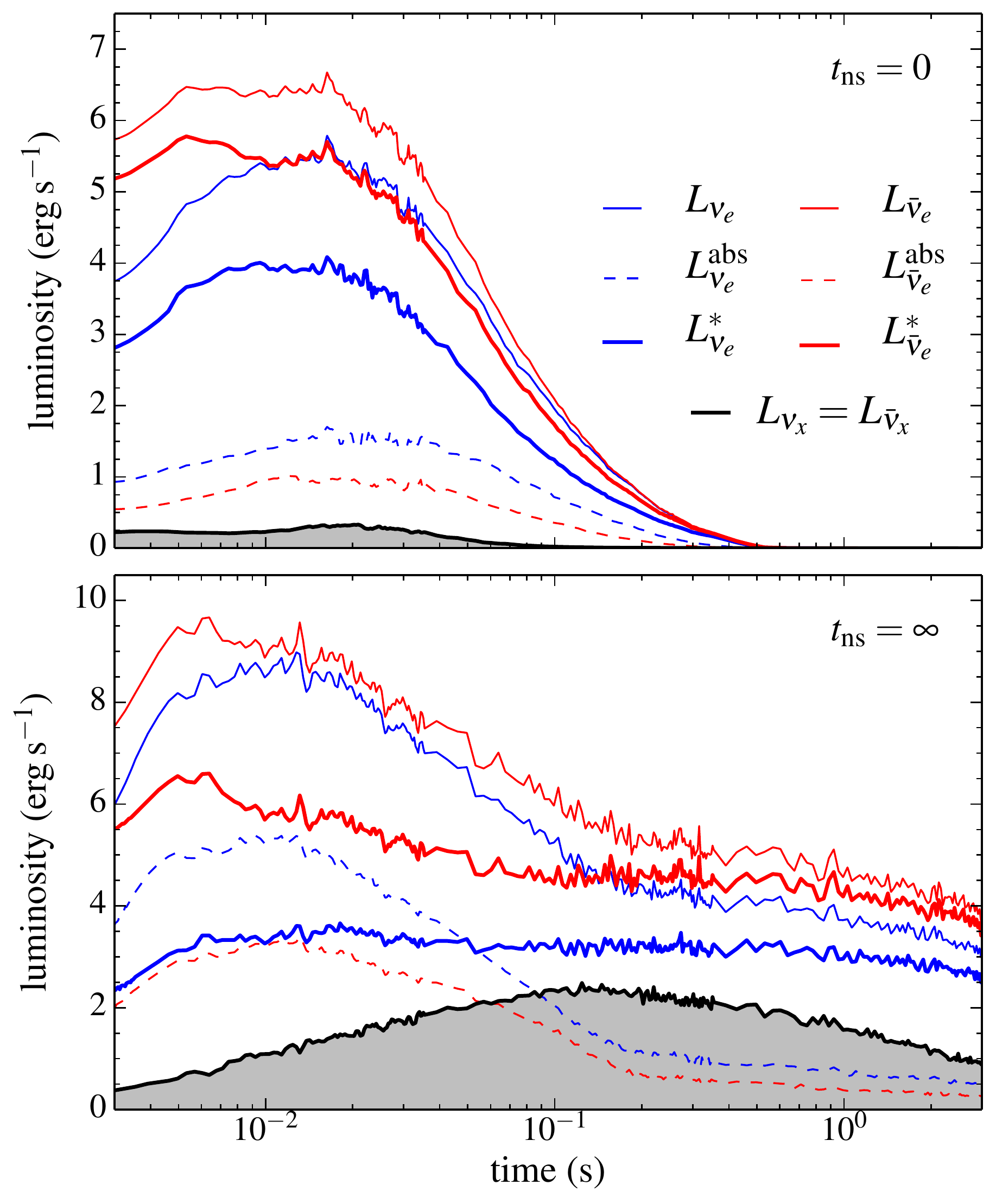}
\caption{Luminosities as a function of time for the prompt BH model
(top) and HMNS model (bottom) without flavor transformation. Shown for electron neutrinos and antineutrinos
are the emitted luminosities $L_{\nu_i}$ (thin solid lines), total power
absorbed $L_{\nu_i}^{\rm abs}$ (dashed lines), and net luminosities used in
equation~(\ref{eq:N_nu}), $L_{\nu_i}^* = L_{\nu_i}-L_{\nu_i}^{\rm abs}$ (thick
solid lines).  The black lines show the emitted luminosities of
heavy lepton neutrinos and antineutrinos, $L_{\nu_x} = L_{\bar{\nu}_x} = L_{\rm
X}/2$ (we neglect their absorption so no correction is applied, i.e., $L_{\nu_x}^* = L_{\nu_x}$).}
\label{fig:lum_abs-corr_time}
\end{figure}

The increase in electron neutrino absorption with flavor transformation
intensity for the long-lived HMNS is the consequence of two effects that modify
the simpler picture for a prompt BH.  First, the drop in electron neutrino
luminosity upon flavor mixing is not as large as in the BH case. The heavy
lepton luminosity is significantly larger than in the pure BH case, as the
boundary layer region reaches higher densities and temperatures
(Figure~\ref{fig:lum_ener_time}).  Also, electron neutrino absorption is more
important than in the pure BH case due to the opaque boundary layer.
Figure~\ref{fig:lum_abs-corr_time} shows that the absorption-corrected
luminosity $L_{\nu_e}^*$ is reduced relative to the emitted luminosity
$L_{\nu_e}$ by a larger factor in model tinf-ab00 than in model BH-ab00 (for
heavy leptons $L^*_{\rm X}=L_{\rm X}$ always, since we neglect their
absorption).  As a result, swapping $L^*_{\nu_x}$ and $L_{\nu_e}^*$ in the HMNS
case results in a moderate (factor $\lesssim 2$) drop in electron neutrino flux
during the relevant part of the evolution, in contrast to the BH case in which
the decrease is a factor $\sim 10$.

The second effect leading to more electron neutrino absorption with flavor
transformation in long-lived HMNS disks is the mixing of the temperature of
emitted neutrinos (Equations~\ref{eq:Teff_nue}-\ref{eq:Teff_nuebar}).  This
effect is present in all models with flavor transformation, and it tends to
increase net absorption by increasing the cross section, as the mean energy of
heavy lepton neutrinos is always larger than that of electron neutrinos or
antineutrinos (Figure~\ref{fig:lum_ener_time}).  For the prompt BH models, this
effect is sub-dominant, since the drop in neutrino flux is much larger than the
increase in mean neutrino energies from the disk (c.f.
Figure~\ref{fig:lum_abs-corr_time}).  For the long-lived HMNS case, however,
the reduction in absorption rate due to the difference between $L_{\nu_e}^*$
and $L_{\nu_x}^*$ is comparable to or smaller than the increase in absorption
rate from the increase in the absorption cross section due to the higher
average neutrino energy. Thus, the global absorption of electron neutrinos
remains nearly constant or even increases. More absorption of electron-type
neutrinos increases the equilibrium electron fraction \cite{Just2022_Yeq}.

\begin{figure}
\includegraphics*[width=\columnwidth]{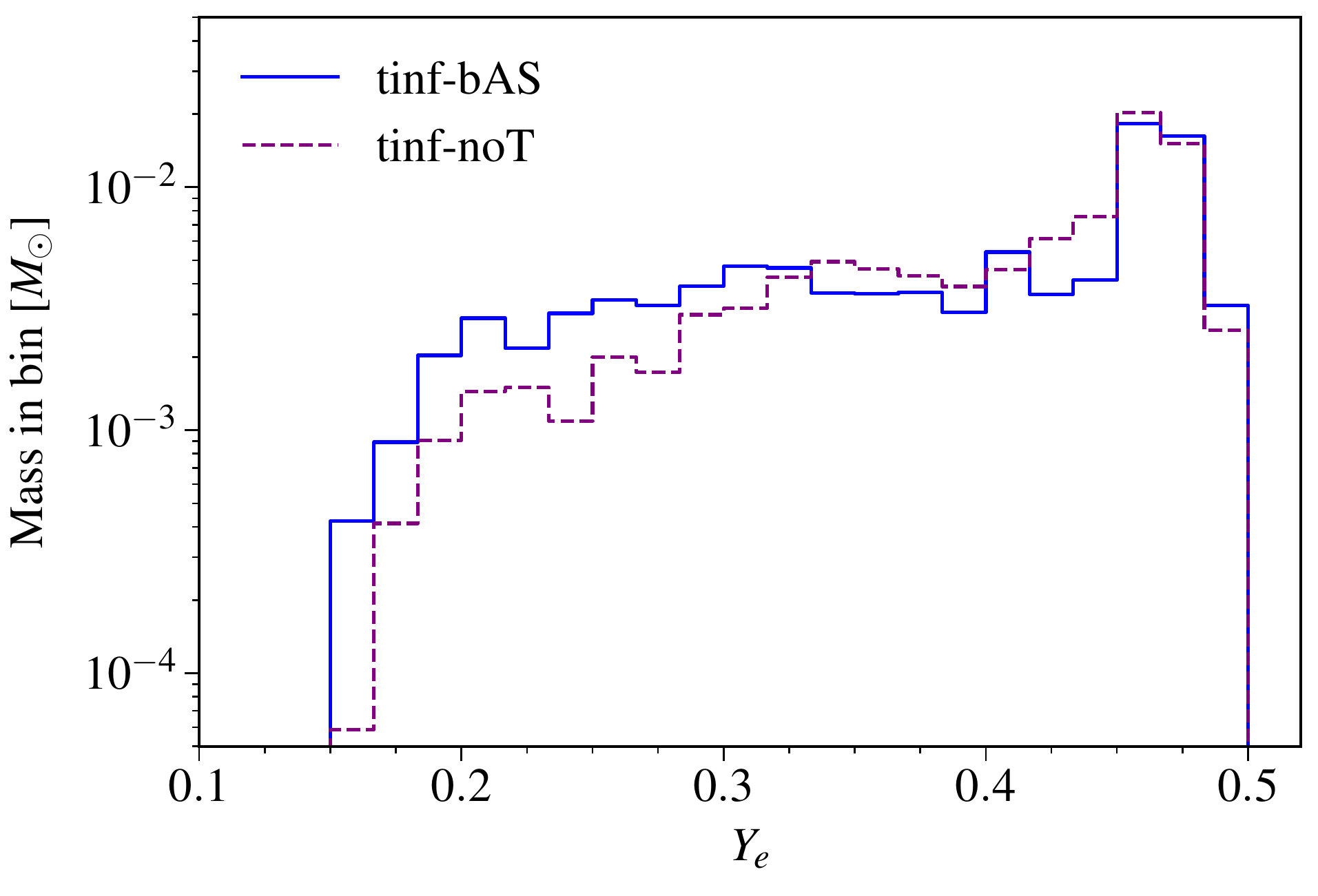}
\caption{Mass histograms of electron fraction for unbound ejecta at the end of
the simulation for a pair of long-lived HMNS models that differ only in that they either
include (tinf-bAS, solid blue) or exclude (tinf-noT, dashed violet) mixing of the neutrino 
temperatures via Equations~(\ref{eq:Teff_nue})-(\ref{eq:Teff_nuebar}). 
}
\label{fig:hist_ye_tinf}
\end{figure}

As a test of this interpretation, we ran another model (tinf-noT) with the same
parameters as tinf-bAS but neglecting the swap of neutrino temperatures. The
absorption of electron neutrinos then decreases during the first
$35$\,ms of evolution relative to model tinf-ab00 and tinf-bAS
(Table~\ref{t:analysis}) as expected, since $L_{\nu_e}^*$ is still
larger than $L_{\nu_x}^*$ by a factor $\geq 2$ over that time period.
The change in $Y_e$ over this interval is also smaller in model tinf-noT than in all
other models with $t_{\rm ns}=\infty$.

Despite the lower amount of electron neutrino absorption in its early evolution
and smaller change in $Y_e$, however, model tinf-noT has a higher average
electron fraction by the end of the simulation (Table~\ref{t:models}) than its
sibling model that includes neutrino temperature oscillation (tinf-bAS).
Figure~\ref{fig:hist_ye_tinf} shows that while the peak of the electron
fraction distribution by the end of the simulation is nearly the same in both
cases, the amount of low $Y_e$ material is lower in model tinf-noT, hence the
average over the entire outflow is higher.

\begin{figure}
\includegraphics*[width=\columnwidth]{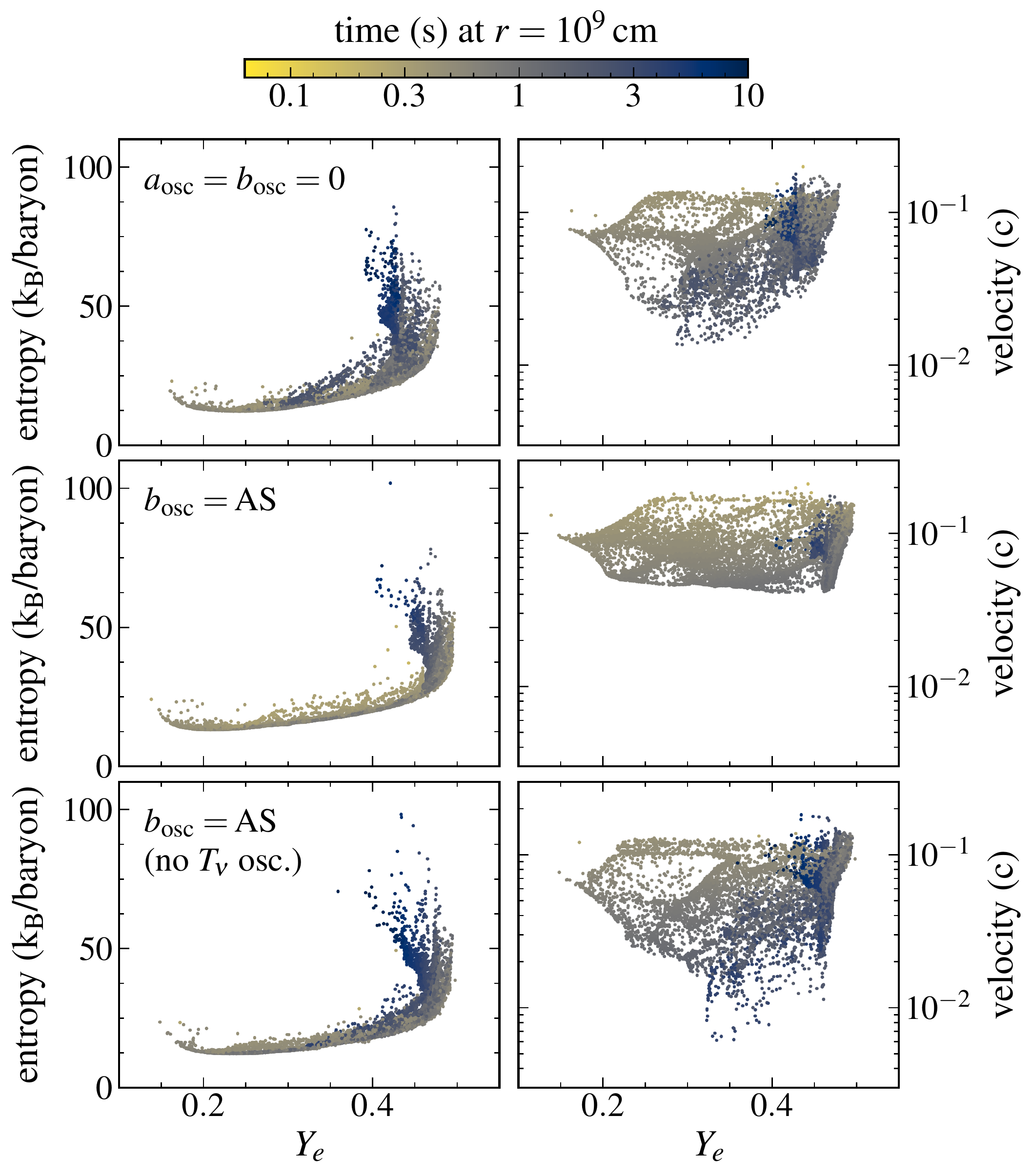}
\caption{Entropy and radial velocity vs. electron fraction for unbound
tracer particles from models with long-lived HMNS and different flavor
transformation configuration: no FFI (tinf-ab00, top), asymmetric (tinf-bAS,
middle) and asymmmetric with no neutrino temperature oscillation (tinf-noT,
bottom). The color shows the time at which the particle last reaches
$r=10^9$\,cm. 
}
\label{fig:ye-ents-vel_scatter}
\end{figure}

To further dissect the origin of these changes, we note that
\citet{lippuner_2017} showed that the outflow from HMNS disks can be separated
into an earlier, mostly neutrino-driven component, and a late component driven
primarily by viscous heating and nuclear recombination. The early component
exhibits a strong correlation between electron fraction and entropy, with a
turnover in the range $Y_e\sim 0.4-0.5$, while the late component shows a
more scattered distribution in entropy in a narrower $Y_e$ range.
Figure~\ref{fig:ye-ents-vel_scatter} shows a scatter plot of unbound particles
in $Y_e$-entropy-velocity space for models tinf-noT, tinf-bAS, and tinf-ab00,
tagged by the time at which they reach the extraction radius at $r=10^9$\,cm.
The presence of the early ($t<1$\,s, yellow) neutrino-driven wind component is
evident, making up the majority of particles that span the electron fraction
interval $[0.15,0.5]$ and forming the broad component of the $Y_e$ histogram in
Figure~\ref{fig:hist_ye_tinf}.  The smaller amount of of low-$Y_e$ ejecta from
model tinf-noT is thus associated with a smaller contribution of the
neutrino-driven wind, given the drop in luminosity upon flavor mixing without
compensation by a higher neutrino temperature.

The late-time component is also evident in Figure~\ref{fig:ye-ents-vel_scatter}
(blue particles), and is associated with the peak in the $Y_e$ histogram. The
fact that this peak is at a similar value of $Y_e$ in models tinf-noT and
tinf-bAS (Figure~\ref{fig:hist_ye_tinf}), but higher than the peak $Y_e$ from model
tinf-ab00 (bottom right panel of Figure~\ref{fig:ye_histogram}), indicates that
its location is much more sensitive to the swapping of fluxes than to that of
neutrino temperatures when the FFI operates.  We can gain a qualitative
understanding of these trends by evaluating the equilibrium electron fraction
from pure absorption, to which a neutrino-driven wind without cooling is driven
\cite{Qian&Woosley96}
\begin{equation}
\label{eq:yeq_abs}
Y_e^{\rm eq,abs} \sim \left(1 + \frac{\langle \varepsilon_{\bar{\nu}_e}\rangle\,L^*_{\bar{\nu}_e}}
                                    {\langle \varepsilon_{\nu_e}\rangle\,      L^*_{\nu_e}} \right)^{-1}
\end{equation}
where again we assume 
$\langle \epsilon_{\nu_i}^2\rangle=\langle \epsilon_{\nu_i}\rangle^2$.
Ignoring attenuation, considering the contribution of the disk alone, and
adopting constant $a_{\rm osc}=b_{\rm osc}=2/3$ we find $Y_e^{\rm eq,abs}\simeq
\{0.35, 0.39, 0.45\}$ at $t=1$\,s for models tinf-ab00, tinf-noT, and tinf-bAS, respectively.
Considering the HMNS contribution alone, we get $Y_e^{\rm eq,abs}\simeq \{0.44,
0.44, 0.48\}$ independent of time for the same set of models. These values are
consistent with model tinf-bAS having a more proton-rich $Y_e$ peak than the
baseline model tinf-ab00, but do not fully account for model tinf-noT being
closer to model tinf-bAS than to model tinf-ab00 in its late-time component.  A
spread in $Y_e$ within a given model can be accounted for by (1) latitude:
particles ejected closer to the rotation axis have a stronger irradiation
contribution from the HMNS, and (2) attenuation: fluctuations in the ratio of
neutron to proton fraction alter the local incident luminosities in
equation~(\ref{eq:yeq_abs}) though the optical depth, and thereby affect
$Y_e^{\rm eq,abs}$.  Also, neutrino emission is non-negligible,
thus a more accurate value of the equilibrium electron fraction would include
all four reactions contributing to the change in electron fraction
(Equations~\ref{eq:gcem}-\ref{eq:ghep}) but is beyond the scope of this study.

Regarding mass ejected and average velocity of the long-lived HMNS outflow,
Table~\ref{t:models} shows that when comparing model tinf-bAS with the
unoscillated model (tinf-ab00), the average outflow velocity increases by $\sim
10\%$ while the mass ejected barely decreases ($\lesssim 2\%$). Removing the
mixing of neutrino temperatures (model tinf-noT) results in a somewhat larger
decrease in ejected mass ($3\%$) and a $5\%$ \emph{decrease} in average
velocity relative to the model without flavor transformation (tinf-ab00).
Figure~\ref{fig:hist_vel} shows the velocity histograms for these models:
flavor transformation without oscillation of the neutrino temperatures produces
more ejecta with low velocities, which is more weakly bound than matter that
expands faster, for similar thermal energy content.  We can attribute this to
the lower absolute amount of absorption given the lower electron neutrino and
antineutrino luminosities. Including temperature mixing increases neutrino
absorption substantially, to the point that the low-velocity tail of the ejecta
distribution is removed. Figure~\ref{fig:ye-ents-vel_scatter} shows that the
missing low-velocity ejecta is primarily late-time, convective outflow that is
more marginally unbound. This removal of low-velocity ejecta is also behind the
trend of increasing average velocity with decreasing ejected mass for $t_{\rm
ns}=\infty$ models with increasing $a_{\rm osc} = b_{\rm osc}$.

Regarding models with finite HMNS lifetime, the set with $t_{\rm ns}=10$\,ms
shows properties similar to the BH set. Tables~\ref{t:models} and
\ref{t:analysis} show that $65\%$ of the $Y_e$ change ($0.11/[\langle
Y_e\rangle-Y_e(t=0)]$) occurs after BH formation for the unoscillated model
(t010-ab00), following the same trend with FFI coefficients as the BH set.  The
same applies to the energy source terms post-BH formation: more viscous heating
and net neutrino cooling, with nearly constant nuclear recombination heating.  

The most notable difference with the prompt BH set is the bump in the electron
fraction histogram at $Y_e \sim 0.1-0.2$ (Figure~\ref{fig:ye_histogram}), which
due to its similarity to models with longer HMNS lifetime, can be attributed to
a more significant neutrino-driven component at early times. This bump decreases in magnitude
and shifts to lower $Y_e$ with increasing FFI coefficients, in line with a
weaker overall neutrino absorption level and a faster decrease of electron
neutrino absorption than antineutrino absorption.

Regarding the model group with $t_{\rm ns}=100$\,ms, Table~\ref{t:analysis}
shows that very little change ($\sim 5\%$) in $Y_e$ occurs after BH formation,
in line with the sharp decrease without recovery of the electron neutrino and
antineutrino luminosities (Figure~\ref{fig:lum_ener_time}). The time integral
of energy source terms also shows a very reduced importance of viscous heating
and neutrino cooling, but a contribution of nuclear recombination that is only
a factor $3$ lower than models for which the earlier phases are also computed
($t_{\rm ns}=0$ and $10$\,ms). The evolution of this set of models is thus
dominated by the earlier HMNS phase, during which neutrino absorption is a
dominant process. 

Comparing the $Y_e$ histograms of this set with those of the $t_{\rm
ns}=\infty$ series (Figure~\ref{fig:ye_histogram}) indicates that a
neutrino-driven wind is clearly present, and becomes stronger with a more
intense FFI.  Figure~\ref{fig:table_results} shows that $t_{\rm ns}=100$\,ms is
the only model set for which the ejected mass increases with more intense
flavor transformation, which we interpret as neutrino absorption taking over as
a driving mechanism of the outflow. We surmise that models with a long-lived
HMNS saturate their mass ejection at nearly $>95\%$ of the initial disk mass,
whereas the model set with $t_{\rm ns}=100$\,ms has room to grow by starting at
$42\%$ of the initial disk mass without FFI effects.

Finally, we find that our results have little sensitivity to the normalization
of the heavy lepton luminosity imposed at the HMNS surface. Models t010-L20,
t100-L20, and tinf-L20 have identical input parameters as the corresponding
asymmetric models t010-bAS, t100-bAS, tinf-bAS, respectively
(Table~\ref{t:models}), except that we set $L_{\rm X,0}^{\rm ns} =
2L_{\nu_e,{\rm 0}}^{\rm ns}$ in equation~(\ref{eq:hmns_lum}).  Comparing each
pair of models with the same $t_{\rm ns}$ in Table~\ref{t:models} shows differences at the few percent
level in all average quantities, with the exception of model t010-L20 which has
an average velocity $10\%$ higher than model t010-bAS, and a mass with $Y_e<0.25$
that is a factor $\sim 2$ higher in the L20 case. 

Table~\ref{t:analysis} shows that for this pair of models (t010-bAS and
t010-L20), the main difference is that electron antineutrino emission after BH
formation is higher in the model with enhanced heavy lepton HMNS luminosity,
with a correspondingly higher net neutrino cooling. Looking at the tinf
counterparts in Table~\ref{t:analysis}, which share the first $10$\,ms of
evolution with the t010 models, we find a $10\%$ higher electron neutrino and
antineutrino absorption in model tinf-L20 than in model tinf-bAS. While the net
change in $Y_e$ is identical in this $35$\,ms HMNS phase, the larger radiative
driving can account for the higher average velocity in model t010-L20 relative
to model t010-bAS. The larger amount of mass with $Y_e < 0.25$ can be
attributed to a more robust neutrino driven wind, which tends to launch more
low-$Y_e$ ejecta at early times (Figure~\ref{fig:ye-ents-vel_scatter}).

We expect the equilibrium $Y_e$ of the long-lived HMNS outflow to have an
important dependency on the imposed electron neutrino and antineutrino
luminosities at the HMNS boundary (normalization and time dependence,
equation~\ref{eq:hmns_lum}), since much of this radiation emitted toward
equatorial latitudes is absorbed at the boundary layer, thus strongly
influencing the $Y_e$ evolution in this region, which acts as 
a reservoir for the outflow. A more extended parameter space study,
or a self-consistent HMNS and disk evolution, would be able to provide
a more physically based characterization of the baseline $Y_e$ of a
long-lived HMNS.

%-----------------------------------------------------------------------------
\subsection{Nucleosynthesis Implications}

\begin{figure*}
\centering
\includegraphics[width=\linewidth]{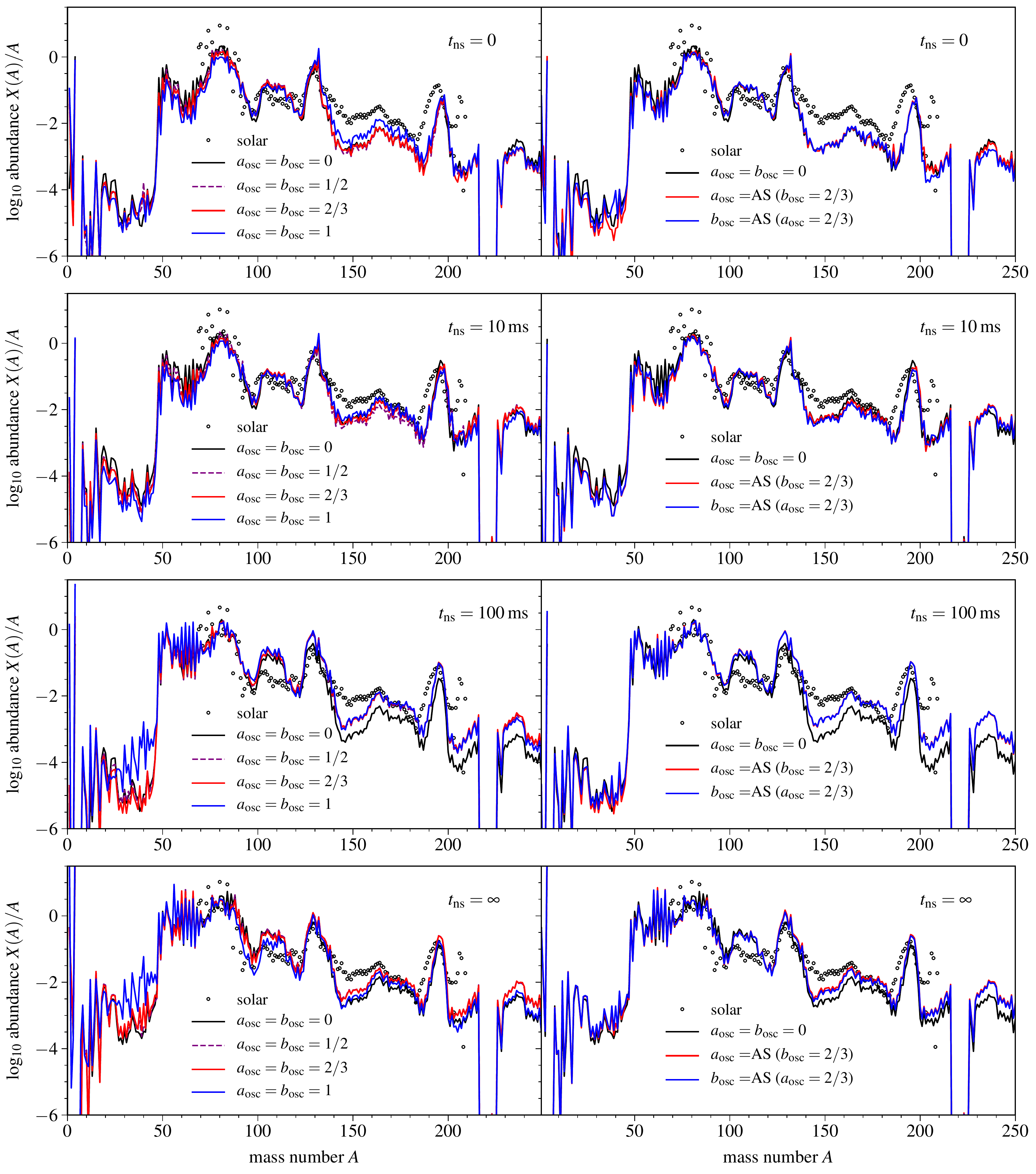}
\caption{Final abundances at $t=30$\,yr as a function of mass number, for
unbound tracer particles from various models. Each row
shows models with different HMNS lifetime, with left and right column showing models
with symmetric and asymmetric flavor transformation coefficients, respectively.
Circles show solar $r$-process abundances from \cite{goriely1999}, scaled to the abundance at
$A=130$ from the model with no flavor transformation ($a_{\rm osc}=b_{\rm osc}=0$), for each
value of $t_{\rm ns}$. Abundances are normalized such that all mass fractions $X(A)$ add up
to unity.}
\label{fig:XvsA_array}
\end{figure*}

Outflows from accretion disks are important contributors to the total
ejecta from NS mergers, an example of which is GW170817, for which the disk ejecta
is expected to have been dominant (e.g.,
\cite{shibata_2017b,metzger_2017}).  The abundance pattern of the ejecta thus
has implications for the $r$-process enrichment contribution (e.g.,
\cite{rosswog_2017,cowan_2019,shibata_2019,siegel_2022}) as well as on the kilonova signal
through the opacities \citep{Kasen+13,tanaka2013,fontes2015} and radioactive
heating rates (e.g., \cite{Li&Paczynski98,Metzger+10b,barnes_2021,klion_2022}). The
$r$-process requires a high abundance of free neutrons when the ejecta
temperature $T\lesssim 5\times 10^9$\,K (e.g., \cite{mendoza_2015}), which
relates directly to the electron fraction of the ejecta shaped by neutrinos at
earlier times when matter is hotter.

Figure~\ref{fig:XvsA_array} shows $r$-process abundances for trajectories from
the same models shown in Figure~\ref{fig:ye_histogram}. Overall, there are no
qualitative changes in the abundance pattern for a given HMNS lifetime,
regardless of the intensity of flavor transformation. More noticeable
differences occur in models with larger $t_{\rm ns}$ due to the increasing
protonization. The general trend is consistent with the $Y_e$ histograms: more
intense flavor transformation produces more heavy $r$-process elements, and
also more light elements (relative to $A\sim 130$) in models with a long-lived HMNS.

For a quantitative assessment, the average mass fraction of lanthanides $X_{\rm
La}$ ($57\leq Z\leq 72$) and actinides $X_{\rm Ac}$ ($89\leq Z\leq 104$) are
shown in Table~\ref{t:analysis} for all models.  Overall, we see that flavor
transformation induces at most a factor $\sim 2$ change in these mass fractions
except for the model set with $t_{\rm ns}=100$\,ms, which shows a larger
variation in the actinide fraction relative to the unoscillated model. A more
significant change of up to a factor of several in $M_\mathrm{ej,red}$ (mass
ejected with $Y_e < 0.25$) is seen in Table~\ref{t:models}, which can alter the
ratio of blue/red kilonovae light curves.

Our models suggest that the FFI introduces quantitative uncertainty in the disk
outflow of at most a factor of two in the mass fraction of heavy $r$-process
elements relevant to kilonova opacities, with minor changes to the overall
$r$-process abundance pattern relative to the standard, no-FFI case.

%-----------------------------------------------------------------------------
\subsection{Comparison with Previous Work}

\citet{Li_Siegel_2021} carried out GRMHD simulations of BH accretion disks
using M1 neutrino transport. Their criterion to activate the FFI stems from a
dispersion relation arising from the linearized evolution equation for neutrino
flavor, with the FFI activated in regions with imaginary frequencies. Once
activated, the FFI manifests as the equality of distribution functions of
neutrinos ($f_{\nu_e} = f_{\nu_\mu} = f_{\nu_\tau}$) and antineutrinos, which
is an equivalent assumption as used in our \emph{Flavor Equilibration} case
($a_{\rm osc} = b_{\rm osc} = 2/3$). In contrast to our models, however, mass
ejection is dominated by magnetic stresses, since it takes several hundred
milliseconds for the disk to reach the radiatively-inefficient stage where the
outflow is driven (mostly) thermally. In this early phase of evolution, the
initial composition of the disk is more important than for late-time outflows
that were fully reprocessed by neutrinos.  Consequently, their un-oscillated
$Y_e$ distribution is much more neutron rich (peaking between $0.15$ and $0.2$)
than that from our un-oscillated prompt BH model
(Figure~\ref{fig:ye_histogram}).

Given that baseline difference, however, introduction of the FFI produces very
similar changes as in our prompt BH simulations.  The $Y_e$ distribution shifts
to more neutron-rich values by $0.01-0.02$, while the unbound mass ejected
decreases by $\sim 10\%$.  While the mass ejection mechanisms are different,
this similarity stems from the fact that their FFI activation criterion results
in widespread operation of the instability, similar to our $\eta_{\rm osc}$
parameter, and flavor swap should alter the emission terms (which dominate in
the pure BH case) in the same way, making the disk more degenerate. Their
$r$-process abundance pattern displays a larger enhancement in heavier elements
when the FFI operates, given the larger relative amounts of ejecta with $Y_e <
0.25$ than in our BH models.
 
\citet{Just2022_FFI} performed axisymmetric viscous hydrodynamic simulations of BH
accretion disks for a time $10$\,s, as well as 3D MHD simulations for a time
$0.5$\,s, with an M1 neutrino scheme. The FFI is activated once the
energy-averaged electron antineutrino flux factor (ratio of number flux to
number density times $c$) exceeds a given value of $0.175$ by default, which
corresponds to a layer below the neutrinosphere where angular asymmetries
relevant to the FFI begin to appear according to a more detailed (static)
analysis.  The neutrinosphere is assumed to be at a flux factor of $1/3$, which
in core-collapse supernovae corresponds to a radial optical depth of $2/3$
\cite{wu_2017_trajectories}. This activation criterion is very similar to our
optical depth based parameter $\eta_{\rm osc}$ (Equation~\ref{eq:eta_osc},
Figure~\ref{fig:eta-osc_snapshots}).  Once active, the FFI is implemented by
algebraically mixing the neutrino number densities and number fluxes of each
flavor, separately for each energy bin of the multi-group M1 scheme. Three
flavor mixing prescriptions are used, among which the assumptions behind their
`mix2' prescription are equivalent to our \emph{Flavor Equilibration} case
($a_{\rm osc} = b_{\rm osc} = 2/3$), while their `mix 1' scheme that conserves
net lepton number shares some similarities with our asymmetric scheme but is
not equivalent.

Their baseline hydrodynamic simulation yields a similar ejecta mass, average
velocity, and electron fraction distributions as our model bh-ab00.
Their model that employs the `mix2' scheme shows a $10\%$ decrease 
in ejecta mass and a decrease of the average electron fraction of $0.02$
relative to their baseline model, which follows the same trend as our
models bh-ab07 compared to bh-ab00 (although we see a larger fractional 
ejecta mass decrease). Similar trends are found in their models
that employ other mixing prescriptions.
Unlike our models, however, the average velocity of all of their
models that include the FFI \emph{decreases} (by $10\%$ for the `mix2' case)
while in our corresponding models the average velocity shows an increase of 
$20\%$ when including the FFI. This discrepancy can be due to the way
in which the average velocity is computed: mass-flux weighted at a fixed
radius in our models (Equation~\ref{eq:vr_ave}) while density-weighted
over a spatial region in theirs. It could also be due to the differences
in absorption resulting from the different neutrino scheme, or 
the implementation of alpha viscosity and how viscous heating reacts
to the increase in degeneracy from more efficient cooling due to the FFI.
Their nucleosynthesis results are entirely consistent with ours.

Our long-lived HMNS model without flavor transformation is in overall
qualitative agreement with that reported in \cite{lippuner_2017}, which used
the same hydrodynamic setup but an older leakage scheme that considered only
charged-current weak interactions, no heavy lepton neutrinos, and did not include an
absorption correction to the disk luminosities. Quantitatively, comparing our
Figure~\ref{fig:ye-ents-vel_scatter} to their Figure~3, the asymptotic $Y_e$ of
their neutrino-driven wind ($\sim 0.55$) is higher than ours ($\sim 0.48$), and
the peak $Y_e$ of their late component ($\sim 0.34$) is lower than what we find
($\sim 0.42$). Both models eject close to $100\%$ of the initial disk mass.

%%%%%%%%%%%%%%%%%%%%%%%%%%%%%%%%%%%%%%%%%%%%%%%%%%%%%%%%%%%%%%%%%%%%%%%
\section{Summary and Discussion \label{s:summary}}

We have studied the effect of the FFI on the long-term outflows from accretion
disks around HMNSs of variable lifetime using axisymmetric, time-dependent,
viscous hydrodynamic simulations. The instability is implemented parametrically
into a 3-species leakage scheme for emission and a disk-lightbulb scheme for
absorption by modifying the absorbed neutrino fluxes and temperatures. We
explore a variety of cases, including partial and complete flavor
equilibration, as well as an ``asymmetric" flavor swap that reflects the
conservation of lepton number in the  neutrino self-interaction Hamiltonian.
Our main results are the following:
\newline

\noindent
1. -- The impact of the FFI on the disk outflow is moderate, changing the total
      unbound mass ejected by up to $\sim 40\%$, the average electron fraction
      by $\sim 10\%$, and in most cases the average velocity by up to $\sim 40\%$
      (Table~\ref{t:models},
      Figure~\ref{fig:table_results}).  The lanthanide and actinide mass fractions of
      the outflow change, in most cases, by up to a factor of $\sim 2$
      (Table~\ref{t:analysis}), with no qualitative changes in the $r$-process abundance
      pattern for a given HMNS lifetime (Figure~\ref{fig:XvsA_array}).  
      \newline

\noindent
2. -- The direction of the changes depends on the HMNS lifetime.
      For a promptly-formed BH or short-lived ($t_{\rm ns}\leq 10$\,ms) HMNS, the
      mass ejected and average electron fraction decrease, and the average velocity
      increases. The composition changes can be traced back to a decrease in the
      electron neutrino/antineutrino absorption with FFI intensity
      (Table~\ref{t:analysis}), which lowers the equilibrium $Y_e$ as well as the
      rate at which this equilibrium is reached (as previously found by
      \cite{Just2022_FFI} for prompt BH disks). The lesser absorption results in
      increased cooling, partially compensated by a higher viscous heating, with the
      net effect of lowering the entropy of the disk. A lower amount of ejected
      material with low velocities 
      accounts for the decrease in mass ejected and
      higher average ejecta velocity (Figure~\ref{fig:hist_vel}).  
      \newline

\noindent
3. -- A longer-lived HMNS ($t_{\rm ns}\geq 100$\,ms) displays a more significant role
	of neutrino absorption in driving the outflow
        (Figure~\ref{fig:ye-ents-vel_scatter}). The FFI results in a more significant
        neutrino driven wind, broadening the electron fraction distribution, increasing
        the peak $Y_e$ to higher values (Figure~\ref{fig:ye_histogram}), increasing the
        average velocity of the ejecta (Figure~\ref{fig:hist_vel}), and increasing the
        mass ejected up to a value of $\sim 95\%$ of the initial disk mass within
        $17.7$\,s of evolution, for a very long-lived HMNS
        (Figure~\ref{fig:table_results}).
	\newline

\noindent
4. -- The trends with HMNS lifetime can be traced back to the effects
      of flavor mixing by the FFI on the neutrino fluxes and temperatures.  For
      BH disks, the heavy lepton luminosity is lower by a factor $\sim 10$ than the
      electron neutrino and antineutrino luminosity, while the mean energies of heavy
      leptons are higher by a factor $\sim 2$ (Figure~\ref{fig:lum_ener_time}). The
      net effect of flavor swap is to decrease absorption (more on electron neutrinos
      than antineutrinos) due to the change in neutrino flux
      (Table~\ref{t:analysis}). For a HMNS disk, on the other hand, the heavy lepton
      luminosity is much higher than for a BH disk  and the amount of electron
      neutrino reabsorption is significant, resulting in a very moderate change in
      the neutrino flux due to the FFI (Figure~\ref{fig:lum_abs-corr_time}). The
      mixing of neutrino temperatures then results in a net \emph{increase} in
      electron neutrino absorption (Table~\ref{t:analysis}), 
      with a protonization of the outflow as well as a more energetic neutrino-driven
      wind that ejects less low-velocity material (Figures~\ref{fig:hist_vel}
      and \ref{fig:ye-ents-vel_scatter}). 
      \newline

\noindent
5. -- Despite the mild changes in composition, the total mass ejected with
      $Y_e < 0.25$ can change by a factor of several (Table~\ref{t:models}),
      thus altering the ratio of red to blue kilonova components if they are to be
      treated separately (e.g., due to spatial segregation).
      \newline

Given the moderate impact of the FFI on the disk outflow, it is natural to
think of this effect as introducing an uncertainty band to theoretical
predictions for the ejecta properties. Our calculations corroborate other 
work \cite{Li_Siegel_2021,Just2022_FFI}, indicating that an overall uncertainty of $\sim 10\%$ in
ejected mass, electron fraction, and velocity, as well as a factor $2$ in
lanthanide/actinide mass fraction can be used as a rule-of-thumb uncertainty in
parameter inference from and/or upper limits on 
multi-messenger observations and galactic abundance studies
(e.g., \cite{kasliwal_2020,thakur_2020,hernandez_2020,wanajo_2021,ricci_2021,holmbeck_2021,geert_2021,chen_2021,gompertz_2022}).  A similar uncertainty level is associated with spatial
resolution of grid-based simulations of post-merger remnants (e.g., \cite{FM13}).  A more difficult
task is to estimate uncertainties in kilonova light curves and spectra due to spatial
segregation of lanthanide-rich vs lanthanide poor material, which would require
radiative transfer simulations to assess the impact on the final outcome (e.g.,
\cite{darbha_2020,Korobkin2021,bulla_2021}).

Our predictions can be made more reliable by (1) improving the quality of
neutrino transport, in particular by using a spectral moment scheme to improve
the angular distribution of radiation for the long-lived HMNS case; (2)
self-consistently including the HMNS-disk system, avoiding the use of separate
luminosities from each object; and (3) including magnetic fields in the
evolution. The latter requires the use of three spatial dimensions, and the
length of time required to fully capture the disk outflow makes such
simulations computationally expensive, precluding an extensive parameter search
with current capabilities. Selected flavor transformation scenarios will need
to be carefully selected for those 3D GRMHD studies to augment the relatively
small number of dynamical models performed to date.

\vspace{0.2in}

\appendix

%---------------------------------------------------------------------
\begin{acknowledgments}

We thank Coleman Dean for comments on the manuscript. We also thank
the anonymous referee for constructive comments that improved the manuscript.
RF, NM, and SF acknolwedge support from the National Sciences and Engineering
Research Council of Canada (NSERC) through Discovery Grants RGPIN-2017-04286
and RGPIN-2022-03463.  SR is supported by a NSF Astronomy \& Astrophysics
Postdoctoral Fellowship under Grant No. 2001760.  We thank the Institute for
Nuclear Theory at the University of Washington for its hospitality and the U.S.
Department of Energy (DOE) for partial support during the completion of this
work.  The software used in this work was in part developed by DOE NNSA-ASC
OASCR Flash Center at the University of Chicago.  This research was enabled in
part by support provided by WestGrid (www.westgrid.ca), the Shared Hierarchical
Academic Research Computing Network (SHARCNET, www.sharcnet.ca), Calcul
Qu\'ebec (www.calculquebec.ca), and Compute Canada (www.computecanada.ca).
Computations were performed on the \emph{Niagara} supercomputer at the SciNet
HPC Consortium \cite{SciNet,Niagara}. SciNet is funded by the Canada Foundation
for Innovation, the Government of Ontario (Ontario Research Fund - Research
Excellence), and by the University of Toronto.  Graphics were developed with
{\tt matplotlib} \cite{hunter2007}.

\end{acknowledgments}

%%%%%%%%%%%%%%%%%%%%

\section{Derivation of the oscillation coefficients for the \emph{Asymmetric} flavor transformation case}
\label{s:asymmetric}

Neglecting collision terms in the quantum kinetic equation, the FFI arises from the neutrino self-interaction Hamiltonian
\begin{equation}
\label{eq:H_nunu}
H_{\nu \nu} = \frac{\sqrt{2}G_{\rm F}}{(2\pi)^3}\int d^3p\, \left( f_\nu - \bar{f}_\nu^*\right)\left(1 - \cos\theta\right)
\end{equation}
where $f_\nu$ is the distribution function of species $\nu$, $\mathbf{p}$ 
is the neutrino momentum, $\theta$ the angle between the direction of the neutrino experiencing
the potential and the momentum in the integrand, 
$G_{\rm F}$ is the Fermi constant, and we have assumed $h=c=1$ (e.g., \cite{richers_2021_pic}).
Equation~(\ref{eq:H_nunu}) satisfies $\bar{H}_{\nu\nu} = -H^*_{\nu\nu}$, which
implies that the probability of flavor transformation is equal for neutrinos and antineutrinos propagating 
in the same direction.
Integrating the quantum kinetic equation over neutrino direction, this symmetry implies conservation of net lepton number, which can be expressed as
\begin{equation}
\label{eq:dist_conservation}
n_{\nu_e} - n_{\nu_x} - n_{\bar{\nu}_e} + n_{\bar{\nu}_x} = \textrm{constant}.
\end{equation}

Defining ``oscillated" number densities as in Equations~(\ref{eq:a_osc})-(\ref{eq:b_osc}), we have
\begin{eqnarray}
n^{\rm osc}_{\nu_e}       & = & (1 - a_{\rm osc})\,n_{\nu_e} + a_{\rm osc}\,n_{\nu_x}\\
n^{\rm osc}_{\nu_x}       & = & (1 - a_{\rm osc})\,n_{\nu_x} + a_{\rm osc}\,n_{\nu_e}\\
n^{\rm osc}_{\bar{\nu}_e} & = & (1 - b_{\rm osc})\,n_{\bar{\nu}_e} + b_{\rm osc}\,n_{\bar{\nu}_x}\\
n^{\rm osc}_{\bar{\nu}_x} & = & (1 - b_{\rm osc})\,n_{\bar{\nu}_x} + b_{\rm osc}\,n_{\bar{\nu}_e}.
\end{eqnarray}
Applying Equation~(\ref{eq:dist_conservation})
\begin{equation}
n^{\rm osc}_{\nu_e} - n^{\rm osc}_{\nu_x} - n^{\rm osc}_{\bar{\nu}_e} + n^{\rm osc}_{\bar{\nu}_x} 
  = n_{\nu_e} - n_{\nu_x} - n_{\bar{\nu}_e} + n_{\bar{\nu}_x},
\end{equation}
and using $n_{\nu_x} = n_{\bar{\nu}_x}$, we obtain
\begin{equation}
a_{\rm osc}\left(n_{\nu_e} - n_{\nu_x}\right) = b_{\rm osc}\left( n_{\bar{\nu}_e} - n_{\bar{\nu}_x}\right),
\end{equation}
which defines the \emph{Asymmetric} flavor transformation case, and is the basis of Equation~(\ref{eq:b_AS}).

\bibliographystyle{apsrev4-2}
\bibliography{ms,references}

\end{document}